\documentclass[prc,aps,nofootinbib,superscriptaddress,showkeys,showpacs,twocolumn,10pt,floatfix]{revtex4-2}
\usepackage{epsfig}
\usepackage{graphicx}
\usepackage[small]{subfigure}
\usepackage{setspace}
\usepackage[T1]{fontenc}
\usepackage[utf8]{inputenc}
\usepackage[english]{babel}
\usepackage{mathrsfs}
\usepackage{braket}
\usepackage[overload]{empheq} 
\usepackage{amssymb}
\usepackage{amsmath}
\usepackage{amsfonts}
\usepackage{siunitx}
\usepackage{amsopn}
\usepackage{mathtools}
\usepackage{float}
\usepackage{chemformula}
\usepackage{comment} 
\usepackage{tikz}
\usepackage{booktabs}
\usepackage{multirow}
\usepackage{verbatim}
\usepackage[normalem]{ulem}
\usepackage{soul}
\usetikzlibrary{arrows}
\usepackage[colorlinks,linkcolor=blue,urlcolor=blue,citecolor=blue]{hyperref}

\usepackage{tikz,xcolor,hyperref}
\definecolor{lime}{HTML}{A6CE39}
\DeclareRobustCommand{\orcidicon}{
	\begin{tikzpicture}
	\draw[lime, fill=lime] (0,0) 
	circle [radius=0.16] 
	node[white] {{\fontfamily{qag}\selectfont \tiny ID}};
	\draw[white, fill=white] (-0.0625,0.095) 
	circle [radius=0.007];
	\end{tikzpicture}
	\hspace{-2mm}
}

\foreach \x in {A, ..., Z}{%
	\expandafter\xdef\csname orcid\x\endcsname{\noexpand\href{https://orcid.org/\csname orcidauthor\x\endcsname}{\noexpand\orcidicon}}
}

\begin{document}

\author{S. Burrello\orcidA}
\email{burrello@lns.infn.it}
\affiliation{INFN, Laboratori Nazionali del Sud, I-$95123$ Catania, Italy}

\author{F. Gulminelli\orcidD}
\affiliation{Université de Caen Normandie, ENSICAEN, CNRS/IN2P3, LPC Caen UMR6534, 14000 Caen, France}

\author{M. Antonelli\orcidE}
\affiliation{Université de Caen Normandie, ENSICAEN, CNRS/IN2P3, LPC Caen UMR6534, 14000 Caen, France}

\author{M. Colonna\orcidB}
\affiliation{INFN, Laboratori Nazionali del Sud, I-$95123$ Catania, Italy}

\author{A. F. Fantina\orcidC}
\affiliation{Grand Accélérateur National d’Ions Lourds (GANIL), CEA/DRF – CNRS/IN2P3, Boulevard Henri Becquerel, 14076 Caen, France}

\title{Bayesian inference of neutron star crust properties \\ using an ab initio-benchmarked meta-model}

\date{\today}

\begin{abstract}
	\noindent 
	Accurate modeling of the neutron star crust is essential for interpreting multimessenger observations and constraining the nuclear equation of state (EoS). However, standard phenomenological EoS models often rely on heuristic extrapolations in the low-density regime, which are inconsistent with microscopic predictions. In this work, we refine a unified meta-modeling framework for the EoS by incorporating low-density corrections based on energy density functionals constrained by ab initio neutron-matter calculations. Using Bayesian inference to combine information from astrophysical observations, nuclear theory, and experiments, we assess the impact of these corrections on key crustal properties, including the crust-core transition density and pressure, crustal composition, and moment of inertia. The improved model reduces uncertainties in the inner crust and emphasizes the importance of low-density physics in EoS modeling, highlighting the value of integrating both theoretical and observational constraints across densities to robustly describe the EoS. Moreover, the adopted approach can be readily applied to any existing EoS model to provide a solid framework for interpreting upcoming high-precision multimessenger data.
\end{abstract}

\maketitle

%\ma{PRC (and many other journals) recommend avoiding the use of "Ref." or "Refs." before citation numbers for brevity and avoiding redundancy. All "Ref." have been removed. Sec. is used for Section.} % remove this disclaimer after reading

\section{Introduction}

Neutron stars (NSs) provide a unique environment to investigate the different phases of hadronic matter~\cite{HaenselBOOK2007,LattimerARNPS2021}. For example, insights into their internal structure have been obtained from recent inferences of their equatorial circumferential radius and gravitational mass~\cite{MillerAJL2019,Miller2021,Riley2019,Riley2021,Salmi2022,Vinciguerra2023,Rutherford2024,Choudhury2024}. Complementary  constraints arise from observations by the LIGO-Virgo-KAGRA collaboration~\cite{AbbottPRL2017,Abbott2019prx,AbbottPRX2023}, with further data expected from upcoming third-generation gravitational-wave (GW) detectors~\cite{MaggioreJCAP2020,Evans2021,BranchesiJCAP2023,Abac2025}. To interpret these observations and extract information on NSs and dense-matter properties, an accurate description of the equation of state (EoS) is needed. Indeed, the relation between the static properties of cold (beta-equilibrated) NSs, relevant for mature isolated objects or coalescing binaries during the inspiral phase, and the underlying microphysics primarily depends on the EoS (see~\cite{HaenselBOOK2007,OertelRMP2017,Burgio2018,BurgioPPNP2021} for a review). This enables the extraction of constraints on the nuclear EoS through analyses combining astrophysical data, nuclear structure, and heavy-ion collision experiments~\cite{HuthNAT2022,TsangNAT2024,KlausnerPRC2025}.

Currently, many inferences of NS properties are based on agnostic EoSs that are not tied to specific microphysical models, e.g.~\cite{EssickPRD2020,GreifMNRAS2019,Landry2019}. This approach is motivated by the fact that macroscopic NS properties like the mass-radius relation are predominantly sensitive to the poorly known high-density core EoS~\cite{AnnalaPRL2018,MillerAJL2019}. Nevertheless, mismatches in modeling the crust and its connection to the core can introduce  systematic errors, highlighting the need for thermodynamically consistent, unified EoS models covering both the crust and the core~\cite{FortinPRC2016,Ferreira2020,Suleiman2021,DavisAA2024}.

It is well established that a transition occurs approximately 1~km beneath the surface from a solid crust to a liquid core. However, the (inner) crust EoS remains affected by considerable model dependence~\cite{HaenselBOOK2007,Chamel2008,Burgio2018}. While these uncertainties are currently small compared to observational precision, they will become increasingly relevant as future GW detectors aim to constrain NS radii at the sub-percent level, supported by improved X-ray measurements~\cite{HuxfordPRD2024}.

Beyond radius measurements, an accurate treatment of crust composition and thickness is also critical for modeling NS thermal evolution~\cite{PagePRL2013,burPRC2015}. Moreover, the crustal fraction of the moment of inertia plays a key role in pulsar glitches, since the crustal superfluid must store sufficient angular momentum to account for the activity observed in the Vela pulsar; see~\cite{amp_review_2023} for a recent review. More precise estimates of the crustal moment of inertia and thickness are therefore required~\cite{SteinerPRC2015}. This motivates current efforts to determine the crust-core (CC) transition point—specifically, the transition density and pressure—using various theoretical models~\cite{VidanaPRC2009,XuAJ2009,MoustakidisPRC2010,burrelloPRC2016,GonzalezPRC2017,LimEPJA2019,GramsEPJA2022}.

To obtain a model-independent probability distribution of the CC transition, recent studies have employed Bayesian techniques within a unified meta-modeling (MM) framework for the nuclear EoS~\cite{CarreauPRC2019,ThiUni2021,DavisAA2024,klausner2025arXiv}. The MM, originally proposed in~\cite{MargueronPRC2018}, has also proven successful in reproducing several NS properties~\cite{MargueronPRC2018II,MontefuscoAA2025}. Within this approach, the response of homogeneous matter to finite-size perturbations offers a qualitative estimate of the CC point~\cite{AnticJPG2019,klausner2025arXiv}. However, the most robust method remains the analysis of phase equilibrium conditions, determining the transition ``from the crust''~\cite{CarreauEPJA2019}. Starting from uncorrelated priors constrained by empirical nuclear data, Bayesian inference identifies the  parameters governing the CC phase transition, incorporating information from both NS physics and microscopic nuclear modeling~\cite{CarreauPRC2019,ThiUni2021,DavisAA2024,klausner2025arXiv}. These studies show that the CC transition is largely insensitive to the high-density EoS, whereas chiral effective field theory ($\chi$-EFT) imposes tighter constraints, particularly via the symmetry-energy slope parameter $L_{\rm sym}$~\cite{LimPRC2019,TewsEPJA2019,EssickPRC2020}. Nonetheless, consistency with \textit{ab initio} neutron-matter pseudo-data at very low densities has often been neglected~\cite{GezerlisPRC2010,CoraggioPRC2013,GandolfiCM2022,PalaniappanPRC2023,PalaniappanPRC2025}, since the MM expansion was not tailored to reproduce pure neutron matter (PNM) in the dilute limit and relies on heuristic extrapolation. This limitation is common among phenomenological energy-density functionals (EDFs), which struggle to reproduce the physics near the unitary limit of an interacting Fermi gas (FG)~\cite{KonigPRL2017,GramsEPJA2022}.

Efforts have been made to bridge the gap between phenomenological and \textit{ab initio} methods~\cite{FurnstahlEPJA2020,BurrelloPLB2020,YangPRC2022}, with the aim of improving EDF predictive power beyond the densities for which they were originally fitted. New EDF classes have been introduced to reconcile constraints from both nuclear saturation and the very-low-density regime~\cite{YangPRC2016,GrassoPPNP2019}. These models, inspired by EFT-based expansions, have been benchmarked against \textit{ab initio} calculations for both PNM and neutron drops. Applications include mean-field studies of finite nuclei and finite-temperature PNM~\cite{burrelloPRC2021,burrelloEPJA2022}.

In this work, we incorporate such parameterizations as a low-density correction blended with the MM approach, ensuring improved consistency with \textit{ab initio} constraints in the sub-saturation regime. We assess the impact on crustal observables, including isotopic composition, crustal moment of inertia, and the CC transition density and pressure. The method, being fully analytical, can be adapted with minimal computational cost to any EoS model to enforce exact low-density behavior. The paper is organized as follows: Sec.~\ref{sec:theo} outlines the theoretical framework, Sec.~\ref{sec:thermo} presents results for the thermodynamical properties of homogeneous and inhomogeneous crust matter, Sec.~\ref{sec:bayesian} is dedicated to the Bayesian analysis and inference of astrophysical observables, and conclusions are drawn in Sec.~\ref{sec:conclusions}.

\section{Theoretical framework}
\label{sec:theo}
\subsection{Meta-model approach}

We consider homogeneous nuclear matter composed of neutrons and protons, characterized by the total baryon number density $n_{\rm B}$ and the isospin asymmetry \mbox{$\delta = (n_{\rm n} - n_{\rm p})/n_{\rm B}$}, where $n_{\rm n}$ and $n_{\rm p}$ are the neutron and proton densities, respectively. The  thermodynamical properties at zero temperature follow from the energy density,
\begin{equation}
	\mathcal{E}_{\rm B}(n_{\rm B}, \delta) = n_{\rm B} \, e_{\rm MM}(n_{\rm B}, \delta) \,,
	\label{eq:epsb}
\end{equation}
where the MM assumes that the energy per baryon $e_{\rm MM}$ is composed of a kinetic contribution $t_{\rm FG}^{\ast}$ and a potential term $v_{\rm MM}$,
\begin{equation}
	e_{\rm MM}(n_{\rm B}, \delta) = t_{\rm FG}^{\ast}(n_{\rm B}, \delta) + v_{\rm MM}(n_{\rm B}, \delta) \,.
	\label{eq:enuc}
\end{equation}
The potential energy $v_{\rm MM}$ in Eq.~\eqref{eq:enuc} is modeled as a polynomial expansion~\cite{MargueronPRC2018},
\begin{equation}
	v_{\rm MM}^{\mathcal{N}} = v_{\rm is}^{\mathcal{N}} + v_{\rm iv}^{\mathcal{N}} = \sum_{\alpha = 0}^{\mathcal{N}} \dfrac{1}{\alpha!} \left( v_{\alpha}^{\rm is} + v_{\alpha}^{\rm iv} \delta^{2} \right) x^{\alpha} \,,
	\label{eq:vn}
\end{equation}
for both the isoscalar $v_{\rm is}$ and isovector $v_{\rm iv}$ components. In Eq.~\eqref{eq:vn}, the expansion parameter
\begin{equation}
	x = \dfrac{n_{\rm B} - n_{\rm sat}}{3n_{\rm sat}} \,,
\end{equation}
is defined in terms of the saturation density $n_{\rm sat}$ of symmetric nuclear matter (SNM). Throughout this work, we adopt $\mathcal{N}=4$~\cite{MargueronPRC2018,DavisAA2024,MontefuscoAA2025}.

Eq.~\eqref{eq:enuc} deviates from a pure parabolic approximation for the isospin dependence because of the kinetic term
\begin{eqnarray}
	t_{\rm FG}^{\ast} =  \dfrac{t_{\rm FG}}{2}  \left[ \left( 1 + \kappa_{\rm sat} \dfrac{n_{\rm B}}{n_{\rm sat}}\right) f_{1} (\delta) + \kappa_{\rm sym} \dfrac{n_{\rm B}}{n_{\rm sat}} f_{2} (\delta) \right]  , \nonumber \\
	\label{eq:tfg_ast}
\end{eqnarray}
where we introduce the kinetic energy per nucleon of SNM
\begin{equation}
	t_{\rm FG} (n_{\rm B}, \delta = 0) = \dfrac{3}{5} \dfrac{\hbar^{2}}{2 m} \left(\dfrac{3\pi^{2}n_{\rm B}}{2}\right)^{2/3} ,
	\label{eq:tfg}
\end{equation}
and the functions
\begin{equation}
	f_{1}(\delta) = \left(1 + \delta\right)^{5/3} + \left(1 - \delta\right)^{5/3} \ ,
	\label{eq:f1}
\end{equation}
\begin{equation}
	f_{2}(\delta) =  \delta \left[ \left(1 + \delta\right)^{5/3} - \left(1 - \delta\right)^{5/3} \right] \ .
\end{equation}
The terms $\kappa_{\rm sat}$ and $\kappa_{\rm sym}$ in Eq.~\eqref{eq:tfg_ast} are linked to the nucleon effective masses 
$m_{\rm q}^{\ast}$ (${\rm q = p, n}$ labels protons and neutrons) by
\begin{eqnarray}
	\kappa_{\rm sat} & = &  \dfrac{m}{m_{\rm sat}^{\ast}} - 1 \qquad \quad \, \, {\rm in \, SNM \,\,} (\delta = 0) \ , \\
	\kappa_{\rm sym} & = & \dfrac{1}{2} \left ( \dfrac{m}{m_{\rm n}^{\ast}} - \dfrac{m}{m_{\rm p}^{\ast}} \right) {\rm \,\, in \,\, PNM} \,\,(\delta = 1) \ ,
\end{eqnarray}
where $m$ is the average nucleon bare mass, $m_{\rm sat}^{\ast} = m^{\ast} (n_{\rm sat},0)$ is the Landau effective mass at saturation~\cite{MargueronPRC2018}, while $\kappa_{\rm sym}$ is directly connected to the isospin effective mass splitting at saturation~$\Delta m^* = m_{\rm n}^{\ast}(n_{\rm sat},1) -  m_{\rm p}^{\ast}(n_{\rm sat},1)$. 
Defining $t_{\rm FG, sat}= t_{\rm FG} (n_{\rm sat},0)$, the model parameters for $\mathcal{N} = 4$ are
\begin{eqnarray}
	v_{0}^{\rm is} & = & E_{\rm sat} - t_{\rm FG, sat} \left( 1 + \kappa_{\rm sat} \right)\\
	v_{1}^{\rm is} & = & - t_{\rm FG, sat} \left( 2 + 5 \kappa_{\rm sat} \right) \\
	v_{2}^{\rm is} & = & K_{\rm sat} - 2 t_{\rm FG, sat} \left( - 1   + 5\kappa_{\rm sat} \right) \\
	v_{3}^{\rm is} & = & Q_{\rm sat} - 2 t_{\rm FG, sat} \left( 4 - 5 \kappa_{\rm sat} \right) \\
	v_{4}^{\rm is} & = & Z_{\rm sat} - 8 t_{\rm FG, sat} \left( -7 + 5 \kappa_{\rm sat} \right)
\end{eqnarray}
for the isoscalar model parameters and 
\begin{eqnarray}
	v_{0}^{\rm iv} & = & E_{\rm sym} - \dfrac{5}{9}t_{\rm FG, sat} \left[ 1 + \left( \kappa_{\rm sat} + 3 \kappa_{\rm sym}\right) \right]\\
	v_{1}^{\rm iv} & = & L_{\rm sym} - \dfrac{5}{9}t_{\rm FG, sat} \left[ 2 + 5 \left( \kappa_{\rm sat} + 3 \kappa_{\rm sym} \right) \right]  \\
	v_{2}^{\rm iv} & = & K_{\rm sym} - \dfrac{10}{9} t_{\rm FG, sat} \left[ - 1 + 5 \left( \kappa_{\rm sat} + \kappa_{\rm sym} \right) \right] \\
	v_{3}^{\rm iv} & = & Q_{\rm sym} - \dfrac{10}{9} t_{\rm FG, sat} \left[ 4 - 5 \left( \kappa_{\rm sat} + 3 \kappa_{\rm sym} \right) \right] \\
	v_{4}^{\rm iv} & = & Z_{\rm sym} - \dfrac{40}{9} t_{\rm FG, sat} \left[ -7 + 5 \left( \kappa_{\rm sat} + 3\kappa_{\rm sym}\right) \right]
\end{eqnarray}
for the isovector counterpart, in terms of the usual empirical parameters $E_{\rm sat}$, $K_{\rm sat}$, $Q_{\rm sat}$, $Z_{\rm sat}$ and $E_{\rm sym}$, $L_{\rm sym}$, $K_{\rm sym}$, $Q_{\rm sym}$, $Z_{\rm sym}$, respectively.
By varying the set of empirical parameters, Eq.~\eqref{eq:vn} defines a parameterized energy for homogeneous matter that enables a convenient exploration of density dependencies, including behaviors not covered by existing Skyrme and relativistic mean-field EDFs.

\subsection{Dilute nuclear matter: YGLO functional }

The YGLO functional developed by \citet{YangPRC2016} incorporates, within the EDF framework, both the explicit density dependence typical of standard effective interactions and the correct behavior of nuclear matter in the dilute regime.

The dilute regime of any Fermi gas is characterized by the condition $|a k_{\rm F}| \ll 1$, where $a$ is the s-wave scattering length and $k_{\rm F} = \left(6\pi^2 n / \nu \right)^{1/3}$ is the Fermi momentum, with $\nu$ denoting the degeneracy and $n$ the number density~\cite{BakerRMP1971}. In this regime, Lee and Yang (LY) proposed in the 1950s an expansion in $(a k_{\rm F})$ for the ground-state energy per particle $e_{\rm LY}$~\cite{LY}, which, in the case where $\nu = 2$, is:
\begin{equation}
	\label{eq:LY_expansion}
	e_{\rm LY} = \epsilon_{\rm F}
	\left[ \frac{3}{5} + \frac{2}{3 \pi} (a k_{\rm F}) + \frac{4  \left ( 11 - 2 \ln 2 \right )}{35 \pi^2} (a k_{\rm F})^2  + \dots \right] \ ,
\end{equation}
where \mbox{$\epsilon_{\rm F} = \hbar^2 k_{\rm F}^2/(2 m)$} is the Fermi energy. % and $\nu = 2$. 
Eq.~\eqref{eq:LY_expansion} has recently been extended up to fourth order within the EFT framework~\cite{WellenhoferPLB2020}. Its truncation to second order remains valid only in the dilute limit, which corresponds to extremely low baryonic densities ($n_{\rm B} \lesssim 10^{-6}$~fm$^{-3}$) in the PNM case due to the large value of the neutron s-wave scattering length $a_{\rm n}$.

To extend the applicability of the expansion to typical nuclear densities, various re-summed expressions have been developed within EFT~\cite{SchaferNPA2005, KaiserNPA2011, KaiserEPJA2012}. A simple re-summed form for the s-wave contribution is given by~\cite{SchaferNPA2005}
\begin{equation}
	e_{\rm resum} = \epsilon_{\rm F} 
	\left[ \frac{3}{5} + \frac{2}{3 \pi} \, \frac{k_{\rm F, n} \, a_{\rm n}}{1 - \frac{6}{35\pi}\left ( 11 - 2 \ln 2 \right )k_{\rm F, n} a_{\rm n} } \right] \ ,
	\label{eq:resummed}
\end{equation}
which also agrees well with recent quantum Monte Carlo (QMC) results~\cite{GezerlisPRC2010}. The resummed expression used here is chosen for its algebraic simplicity. Other resummation strategies, including effective-range corrections~\cite{KaiserEPJA2012}, difermion approaches~\cite{SchwenkPRL2005}, and Padé approximants~\cite{WellenhoferPRR2020}, would require additional inputs beyond the minimal parameter set adopted in this study.

The YGLO functional is constructed to reproduce the first two terms of the potential part of the LY expansion given in Eq.~\eqref{eq:LY_expansion}. The corresponding potential energy density for SNM  (i = s) and PNM (i = n) reads
\begin{equation}
	\label{eq:yglo}
	\mathcal{V}_{\rm i}^{\rm Y} = Y_{\rm i} [n_{\rm B}] n_{\rm B}^2 + D_{\rm i} n_{\rm B}^{8/3} + F_{\rm i} n_{\rm B}^{\alpha + 2} \ ,
\end{equation}
where the re-summed term $Y_{\rm i}[n_{\rm B}]$ is defined as
\begin{equation}
	Y_{\rm i} [n_{\rm B}] = \frac{B_{\rm i}}{1 - R_{\rm i} n_{\rm B}^{1/3} + C_{\rm i} n_{\rm B}^{2/3}} \ ,
\end{equation}
and the coefficients $B_{\rm i}$ and $R_{\rm i}$ are constrained by matching to the LY expansion up to second order in $(a_{\rm i} k_{\rm F,i})$,
\begin{align}
	\label{eq:bi_ri}
	B_{\rm i} & = \frac{2 \pi \hbar^2}{m} \frac{\nu_{\rm i} - 1}{\nu_{\rm i}} a_{\rm i} \ , \nonumber                             \\ 
	R_{\rm i} & = \frac{6}{35 \pi} \left ( \frac{6 \pi^2}{\nu_{\rm i}} \right )^{1/3} \left ( 11 - 2 \ln 2 \right ) a_{\rm i} \ , 
\end{align}
where we adopt $a_{\rm i} = -18.9$~fm ($-20.0$~fm) and degeneracy $\nu_{\rm i} = 2$ (4) for PNM (SNM), respectively. The parameters $D_{\rm i}$, $F_{\rm i}$, and $C_{\rm i}$ were determined in~\cite{YangPRC2016} by fitting QMC AV4 pseudodata~\cite{GezerlisPRC2010} for $n_{\rm B} \le 0.005$~fm$^{-3}$, and two benchmark $\chi$-EFT calculations at higher densities: Akmal~\cite{AkmalPRC1998} and FP~\cite{FriedmanNPA1981}. In all cases, $\alpha = 0.7$. This resulted in two parameter sets, labeled YGLO (Akmal) and YGLO (FP), reported in Table~\ref{tab:yglo_parameters}.

In this work, we introduce the new parameterization YGLO (MU), fitted to recent third-order PNM Bogoliubov many-body perturbation theory (BMBPT3) calculations based on renormalization group evolved low-momentum interactions, including three-nucleon (3N) induced terms~\cite{PalaniappanPRC2023,PalaniappanPRC2025}. These results show excellent agreement with other recent \textit{ab initio} methods, namely Brueckner–Hartree–Fock calculations of~\cite{VidanaFRO2021} up to $k_{\rm F, n} = 0.53$~fm$^{-1}$ ($n_{\rm B} \approx 0.005$~fm$^{-3}$), AFDMC PNM data~\cite{GandolfiCM2022} up to $k_{\rm F, n} = 0.8$~fm$^{-1}$ ($n_{\rm B} \approx 0.02$~fm$^{-3}$) and MBPT3 results~\cite{HebelerPR2021} up to $k_{\rm F, n} = 1.35$~fm$^{-1}$ ($n_{\rm B} \approx 0.08$~fm$^{-3}$). At higher densities, contributions from the bare 3N interaction and higher two-nucleon (2N) partial waves become increasingly important, limiting the reliability of purely 2N-based predictions. The value $\alpha = 0.7$ is retained from the original YGLO functional introduced in Ref.~\cite{YangPRC2016}. Fixing $\alpha$ preserves the original SNM EoS and avoids introducing additional free parameters in the present YGLO (MU) parameterization. However, this choice might limit the flexibility of the YGLO (MU) parameterization for applications beyond the specific domain considered in this work.

The parameters in Table~\ref{tab:yglo_parameters}, with their different sign and magnitude, reflect the density expansion structure of the functional. We have verified the stability of the fit by varying the density range up to \(n_{\rm B} = 0.08\,\text{fm}^{-3}\). However, below \(n_{\rm B} = 0.02\,\text{fm}^{-3}\), the results are nearly model-independent, being only mildly influenced by the parameter \(C_{\rm n}\), whose amplitude is reflected in the behavior of the different parameterizations.

\begin{table}[h]
	\centering
	\begin{tabular}{ccccccc}
		\toprule
		YGLO  & $C_{\rm s}$ & $D_{\rm s}$ & $F_{\rm s}$ & $C_{\rm n}$ & $D_{\rm n}$ & $F_{\rm n}$ \\
		\midrule
		Akmal & 8.188       & -6624.87    & 6995.46     & 70.19       & -8377.83    & 8743.85     \\
		FP    & 8.188       & -6624.87    & 6995.46     & 100.87      & -9264.18    & 9571.90     \\
		MU    & 8.188       & -6624.87    & 6995.46     & 90.87       & -9427.83    & 9706.90     \\
		\bottomrule
	\end{tabular}%
	\caption{Values of the adjusted parameters obtained for three parameterizations of YGLO, differing in their PNM density behavior. The units considered in the table are fm$^2$, MeV fm$^5$, and MeV fm$^{3 + 3\alpha}$, for $C_{\rm i}$, $D_{\rm i}$, and $F_{\rm i}$ (i = s, n), respectively.}
	\label{tab:yglo_parameters} 
\end{table}

\begin{figure} [tbp!]
	\includegraphics[width=\linewidth]{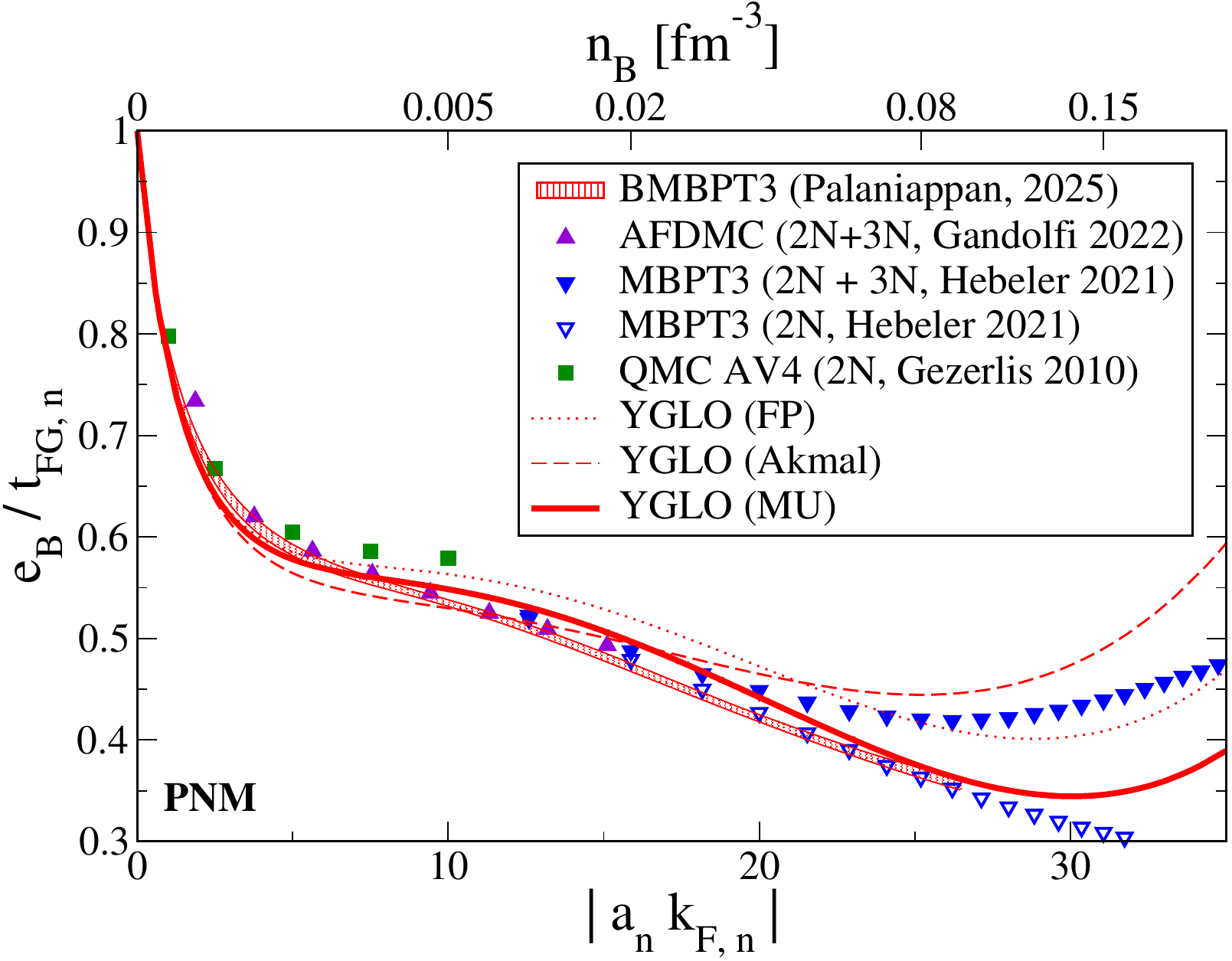}
	\caption{Energy per nucleon in PNM, normalized to the free Fermi gas value $(e_{\rm B} / t_{\rm FG,n})$, as a function of $(a_{\rm n}k_{\rm F,n})$, for the three YGLO parameterizations. Shown for comparison are various available \textit{ab initio} results: QMC AV4 pseudodata from~\cite{GezerlisPRC2010} (green squares), AFDMC results from~\cite{GandolfiCM2022} (violet upward triangles), MBPT3 calculations from~\cite{HebelerPR2021} (blue downward triangles), and BMBPT3 predictions from~\cite{PalaniappanPRC2025} (red-striped band). Selected values of $n_{\rm B}$ are indicated on the top horizontal axis.}
	\label{fig:e_efg_abinitio_YGLO_Akmal_FP_MU}
\end{figure}

Following~\cite{burrelloPRC2021}, the YGLO functional can be extended to arbitrary isospin asymmetry $\delta$, and expressed as the sum of a kinetic term $t_{\rm FG}$ and a potential contribution $v_{\rm Y}$:
\begin{eqnarray}
	e_{\rm Y}(n_{\rm B}, \delta) &=& t_{\rm FG}(n_{\rm B}, \delta) + v_{\rm Y}(n_{\rm B}, \delta) \,,
	\label{eq:eyglo}
\end{eqnarray}
with the potential part given by
\begin{eqnarray}
	v_{\rm Y}(n_{\rm B}, \delta) = \dfrac{1}{n_{\rm B}} \left[ \mathcal{V}_{\rm s}^{\rm Y} + \left(\mathcal{V}_{\rm n}^{\rm Y} - \mathcal{V}_{\rm s}^{\rm Y} \right) \delta^2 \right] \,,
	\label{eq:vyglo}
\end{eqnarray}
and the kinetic energy per nucleon
\begin{eqnarray}
	t_{\rm FG}(n_{\rm B}, \delta) = \dfrac{1}{2} t_{\rm FG}(n_{\rm B}, 0) f_1(\delta) \,,
\end{eqnarray}
since $\kappa_{\rm sat} = \kappa_{\rm sym} = 0$ in the YGLO functional.

Figure~\ref{fig:e_efg_abinitio_YGLO_Akmal_FP_MU} displays the energy per nucleon $e_{\rm B}$ in PNM, normalized to the free Fermi gas energy,
\begin{equation}
	t_{\rm FG, n} = \frac{3}{5} \frac{\hbar^2 k_{\rm F, n}^2}{2 m_{\rm n}} \,,
	\label{eq:e_Fg}
\end{equation}
as a function of $(a_{\rm n} k_{\rm F, n})$. The results from the three YGLO parameterizations listed in Table~\ref{tab:yglo_parameters} are shown together with benchmark \textit{ab initio} calculations~\cite{GezerlisPRC2010,HebelerPR2021,GandolfiCM2022,PalaniappanPRC2025}. It is worthwhile to note that all YGLO variants reproduce the density dependence of the PNM energy within the $\chi$-EFT band from the dilute limit up to saturation~\cite{YangPRC2016}. Given that the three parameterizations yield similar results for $n_{\rm B} \lesssim 0.10$~fm$^{-3}$, we adopt the YGLO (MU) set—based on the most recent and complete \textit{ab initio} input—and hereafter refer to it simply as YGLO, unless stated otherwise.

\subsection{\textit{Ab initio} benchmarked correction at sub-saturation densities}

The MM is not designed to accurately reproduce the low-density behavior of PNM as predicted by the LY expansion and \textit{ab initio} results.\footnote{
    In the MM an \textit{ad hoc} exponential correction was introduced to suppress the potential energy in the zero-density limit~\cite{MargueronPRC2018}. In later works~\cite{AnticJPG2019,Rahul2021}, the parameters of this correction were adjusted to improve performance, but the quality of the parametrization was never tested below $n_{\rm B} \approx 0.02$~fm$^{-3}$.}

\begin{figure*}[tbp!]
	\includegraphics[width=0.48\linewidth]{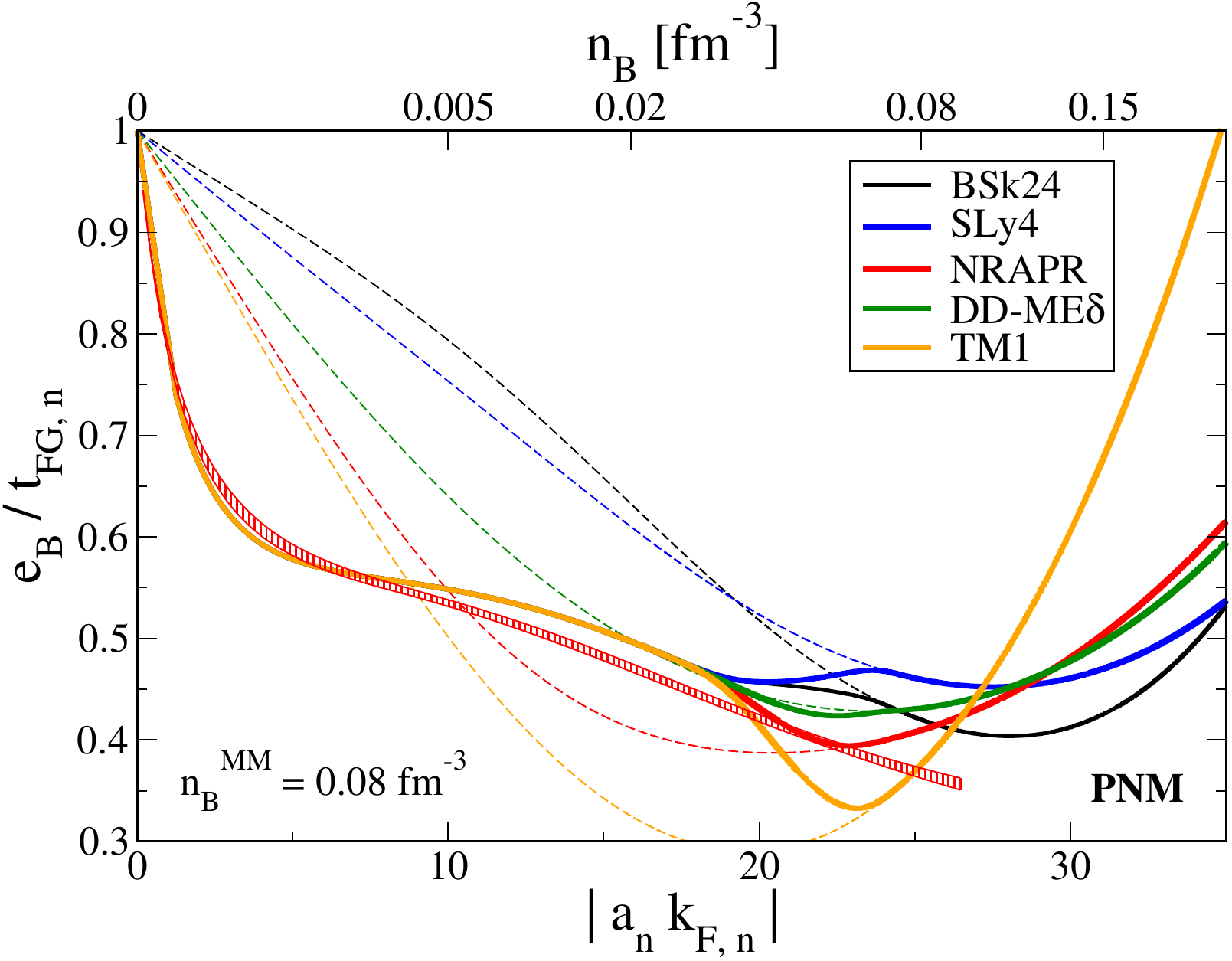} \quad \includegraphics[width=0.48\linewidth]{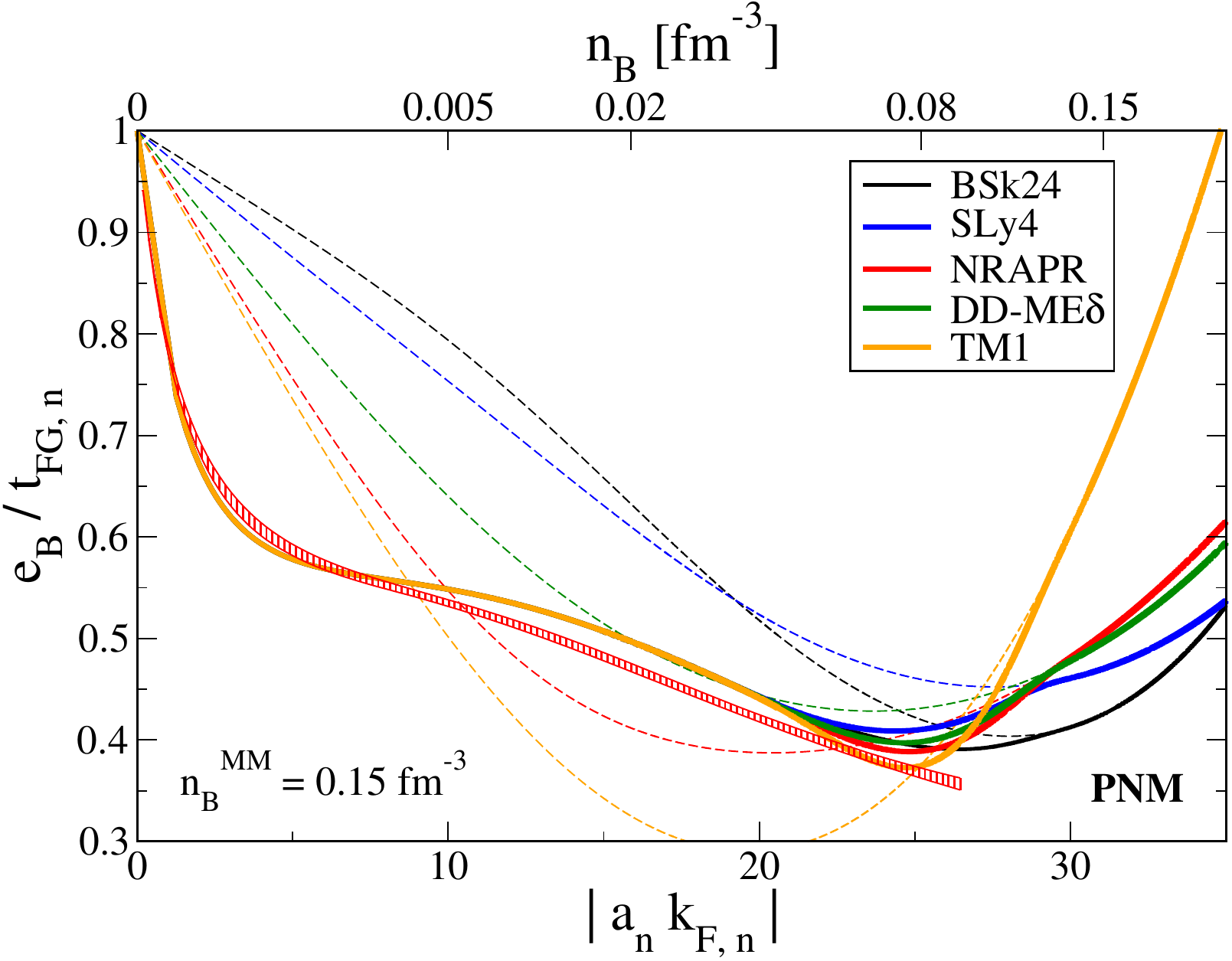}
	\caption{Energy per nucleon in PNM, normalized to the free Fermi gas energy $(e_{\rm B} / t_{\rm FG, n})$, as a function of $(a_{\rm n}k_{\rm F, n})$, obtained with the standard MM (thin dashed lines), or by using YGLO below $n_{\rm B}^{\chi} = 0.02$~fm$^{-3}$ and blending in the range $\left[ n_{\rm B}^{\chi}, n_{\rm B}^{\rm MM} \right]$ (thick solid lines). Two values of $n_{\rm B}^{\rm MM}$ are considered: 0.08~fm$^{-3}$ (left) and 0.15~fm$^{-3}$ (right). Various empirical EDF parameter sets are explored. The red-striped band shows the benchmark \textit{ab initio} calculations from~\cite{PalaniappanPRC2025}.}
	\label{fig:e_efg_abinitio_MM_MU}
\end{figure*}

Our goal is to incorporate a suitable correction into the MM approach to improve its agreement with nuclear theory constraints in the sub-saturation density regime. To this end, we enforce that the energy per baryon $e_{\rm B}$ matches the YGLO prediction $e_{\rm Y}$ for $n_{\rm B} \le n_{\rm B}^{\chi} = 0.02$~fm$^{-3}$. This value corresponds to the lowest density for which $\chi$-EFT predictions used in this work are available and marks the onset of significant three-body effects, where model dependence becomes relevant in \textit{ab initio} calculations~\cite{HebelerPR2021,PalaniappanPRC2025}.

%At higher densities, constraints from MBPT calculations with $\chi$-EFT interactions remain valuable~\cite{HuthNAT2022}, but uncertainties increase. Thus, the energy must blend into an empirical description like the MM approach, which includes complementary information from independent sources such as nuclear phenomenology. 
At higher densities, MBPT calculations based on $\chi$-EFT interactions still offer valuable guidance, although uncertainties grow due to the breakdown of the low-momentum expansion. For practical purposes, the energy of homogeneous matter is then smoothly connected to an empirical description, such as the MM approach, which incorporates complementary constraints from nuclear phenomenology. This blending is carried out up to a final density $n_{\rm B}^{\rm MM}$, beyond which the standard MM expansion is fully recovered. In the intermediate region $n_{\rm B}^{\chi} \le n_{\rm B} \le n_{\rm B}^{\rm MM}$, a smooth blending procedure is implemented:
\begin{equation}
	e_{\rm B}(n_{\rm B}, \delta) = e_{\rm Y}(n_{\rm B}, \delta) \left(1 - \eta_{\chi}^{\rm MM} \right) + e_{\rm MM}(n_{\rm B}, \delta) \eta_{\chi}^{\rm MM} \,,
	\label{eq:interpolation}
\end{equation}
where $\eta_{\chi}^{\rm MM}$ is a smooth transition function,
\begin{equation}
	\eta_{\chi}^{\rm MM}(n_{\rm B}) = \dfrac{f(x_{\chi}^{\rm MM})}{f(x_{\chi}^{\rm MM}) + f(1 - x_{\chi}^{\rm MM})} \,,
	\label{eq:mollifier}
\end{equation}
defined through
\begin{equation}
	f(x_{\chi}^{\rm MM}) = 
	\begin{cases}
		e^{-1/x_{\chi}^{\rm MM}} , & x_{\chi}^{\rm MM} > 0 \\
		0 ,                        & x_{\chi}^{\rm MM} \le 0
	\end{cases}
\end{equation}
and
\begin{equation}
	x_{\chi}^{\rm MM} = \dfrac{n_{\rm B} - n_{\rm B}^{\chi}}{n_{\rm B}^{\rm MM} - n_{\rm B}^{\chi}} \,.
\end{equation}
The transition function $\eta_{\chi}^{\rm MM} (n_{\rm B})$ above is infinitely smooth over the whole real line, thus ensuring the continuity of the energy per baryon $e_{\rm B} (n_{\rm B}, \delta)$ and all its density derivatives. Exploring other functional forms and quantifying the impact on the crustal properties will be pursued in a future work, as it goes beyond the scope of the present study.

\subsection{Homogeneous nuclear-matter properties}

The blending procedure ensures the thermodynamic continuity for the homogeneous bulk matter. Once the energy density is defined, we can easily obtain all the other zero-temperature thermodynamic quantities. For each species ${\rm q}$, we have (our chemical potentials do not account for rest mass): 
\begin{equation}
	\mu_{\rm q} (n_{\rm B}, \delta) =  \left( \dfrac{\partial \varepsilon_{\rm B}}{\partial n_{\rm q}}\right)_{\rm n_{\rm q^{\prime}}} = e_{\rm B} + n_{\rm B} \left( \dfrac{\partial e_{\rm B}}{\partial n_{\rm q}}\right)_{\rm n_{\rm q^{\prime}}} \ ,
	\label{eq:muq1}
\end{equation}
where the partial derivatives with respect to the density $n_{\rm q}$ of particles of type ${\rm q}$ are computed keeping the density of the other species $n_{\rm q^{\prime}}$ constant.
Likewise, in terms of $x$ and $\delta$:
\begin{equation}
	\mu_{\rm q} (n_{\rm B}, \delta) = e_{\rm B} + \dfrac{1 + 3x}{3} \left( \dfrac{\partial e_{\rm B}}{\partial x}\right)_\delta + \left( \tau_{3} - \delta \right)\left( \dfrac{\partial e_{\rm B}}{\partial \delta}\right)_x \ .
	\label{eq:muq}
\end{equation}
Finally, the nuclear pressure can be calculated from (the derivatives of $e_{\rm B}$ are given in Appendix~\ref{app:derivatives_fg})
\begin{equation}
	P_{\rm B} (n_{\rm B}, \delta) =  \sum_{\rm q = n,p} n_{\rm q}\mu_{\rm q} - n_{\rm B} e_{\rm B} (n_{\rm B}, \delta) \ .  
\end{equation}
Whatever the transition function $\eta_{\chi}^{\rm MM} (n_{\rm B})$, the approach in Eq.~\eqref{eq:interpolation} may cause the emergence of spurious instabilities (negative values of the chemical-potential derivative in the PNM case). Hence, it is necessary to constrain the transition function $\eta_{\chi}^{\rm MM} (n_{\rm B})$ by demanding that 
\begin{equation}
	\dfrac{\partial \mu_{\rm n}}{\partial n_{\rm B}} \ge 0    
\end{equation}
everywhere for the PNM. Then, according to Eq.~\eqref{eq:muq1}, 
\begin{eqnarray}
	\dfrac{\partial^{2} e_{\rm B}}{\partial n_{\rm B}^{2}} + \dfrac{2}{n_{\rm B}} \dfrac{\partial e_{\rm B}}{\partial n_{\rm B}} \ge 0 \ .
	\label{eq:inequality}
\end{eqnarray}
The inequality expressed by Eq.~\eqref{eq:inequality} is detailed in Appendix~\ref{app:inequality}. For its practical implementation, we evaluate Eq.~\eqref{eq:inequality} on a dense grid of densities up to $n_{\rm B}^{\rm MM}$; parameter sets that violate the condition at any point are discarded.

\subsection{Inhomogeneous matter in the crust}
\label{sec:inhomogeneous}

To model the crust, we consider Wigner-Seitz (WS) cells immersed in a homogeneous electron gas, each containing a single nucleus with mass number $A$ and proton number $Z$ at its center. The volume of a WS cell, $V_{\text{WS}}$, is determined to ensure charge neutrality:
\begin{equation}
	V_{\text{WS}} = \frac{Z}{n_{\rm p}}, \label{eq:vws}
\end{equation}
where $n_{\rm p}$ represents the total proton density in each cell. 
Once the neutron-drip conditions are reached in the crust, neutrons begin to drip off nuclei, thus clusters are immersed in a background of unbound neutrons. 
This regime cannot be reproduced in terrestrial experiments. As a result, ground-state properties of the inner crust must be determined entirely through theoretical modeling. Uncertainties therefore stem not only from the treatment of homogeneous matter, but also from the many-body method adopted for describing clusterized systems.

We model crustal inhomogeneities using the compressible liquid drop model (CLDM) (see e.g. Refs.~\cite{BaymNPA1971,DouchinAA2001}), whose predictions qualitatively agree with more microscopic approaches like (extended) Thomas-Fermi methods~\cite{CarreauAA2020,Grams2022,klausner2025arXiv,Grams2025}. 
Moreover, its computational efficiency makes it well suited for Bayesian analyses, while maintaining a unified crust-core treatment, meaning that the same functional is used for both the core and the bulk part of the crust. Notably, the CLDM allows one to disentangle the physical contributions to the cluster binding energy $E_{\rm ion}$, which can be written as
\begin{eqnarray}
	E_{\rm ion} &=& (A - Z)m_{\rm n} c^2 + Z m_{\rm p} c^2 \nonumber \\
	&+& E_{\rm bulk} + E_{\rm Coul} + E_{\rm surf + curv},
\end{eqnarray}
where $E_{\rm bulk} = A e_{\rm B}(n_{\rm ion}, I)$, and $e_{\rm B}(n_{\rm ion}, I)$ is the energy per baryon of homogeneous matter at density $n_{\rm ion}$ and isospin asymmetry~$I = 1 - 2Z/A$.

For simplicity, we restrict our analysis to spherical nuclear clusters. The impact of non-spherical structures, the so-called `pasta phases', was already discussed in Refs.~\cite{ThiEPJA2021, ThiAA2021} and shown to have a negligible impact on the EoS and not to substantially modify the CC transition point.
Under this assumption, the Coulomb energy is expressed as \cite{Ravenhall1983b}:
\begin{eqnarray}
	E_{\mathrm{Coul}} &=& \frac{2}{5} \pi (e n_{\rm ion} r_{\rm N})^2 \phi \left(\dfrac{Z}{A}\right)^2 \nonumber \\
	&\times& \left[\phi + 2 \left(1 - \frac{3}{2} \phi^{1/3}\right)\right] V_{\rm WS} \ ,
	\label{eq:Ecoul}
\end{eqnarray}
where $e$ is the elementary charge, $\phi = A / (n_{\rm ion} V_{\rm WS})$ is the
volume fraction of the cluster, and $ r_{\rm N} = \left( \dfrac{3A}{4\pi n_{\rm ion}} \right)^{1/3} $ denotes the nuclear radius\footnote{
    The last term of Eq.~\eqref{eq:Ecoul} corresponds to the lattice energy, that can be equivalently expressed as $-0.9 e^2 Z^2 \phi^{1/3} /(r_N V_{\rm WS})$. In this work, we used the Madelung constant $\approx 0.896$ \cite{HaenselBOOK2007} instead of the coefficient $0.9$.}.

Calculating the surface energy from the energy functional requires strong approximations, to introduce gradient terms with additional phenomenological parameters~\cite{NikolovPRC2011}. However, values for the surface tension, defined as $\sigma_{\rm s} (A, I) = E_{\rm surf} / S$, with $S$ being the nuclear surface, remains model dependent, especially at extreme $I$ values in the inner crust. Moreover, the presence of a surrounding neutron gas further modifies the in-medium surface energy. 
Similar considerations hold also for the curvature tension $\sigma_{\rm c}$. Consequently, parametrized expressions are typically employed. 
We adopt the same functional form as in Refs.~\cite{MaruyamaPRC2005,NewtonAAS2013}, that are: %\aff{[AFF: I think $n_{ion}$ goes in the denominator, please check]}
\begin{equation}
	E_{\text{surf} + \text{curv}} = \dfrac{3A }{r_{\rm N}n_{\rm ion}} \left[ \sigma_{\rm s}(I) +\dfrac{ 2\sigma_{\rm c}(I)}{r_{\rm N}} \right] \ , %AFF I have moved n_ion in the denominator
%	E_{\text{surf} + \text{curv}} = \dfrac{3A n_{\rm ion}}{r_{\rm N}} \left[ \sigma_{\rm s}(I) +\dfrac{ 2\sigma_{\rm c}(I)}{r_{\rm N}} \right] \ , 
\end{equation}
where $ \sigma_{\rm s} $ and $ \sigma_{\rm c} $ are defined as~\cite{RavenhallNPA1983}
\begin{equation}
	\sigma_{\rm s} (I) = \sigma_0 \frac{2^{4} + b_{\rm s}}{y_{\rm p}^{-3} + b_{\rm s} + (1 - y_{\rm p})^{-3}} \ ,
	\label{eq:sigmas}
\end{equation}
\begin{equation}
	\sigma_{\rm c} (I) = 5.5 \sigma_{\rm s} (I) \frac{\sigma_{0, {\rm c}}}{\sigma_0} (\beta - y_{\rm p}) \ ,
\end{equation}
where $ y_{\rm p} = (1 - I)/2 $, and the surface parameters $ (\sigma_0, \sigma_{0, {\rm c}}, b_{\rm s}, \beta) $ are properly optimized for each set of bulk parameters, see Sec.~\ref{sec:bayesian}. 

For a given baryonic density $n_{\rm B}$, the total energy density $\mathcal{E}_{\rm WS}$ in each WS cell can be written as:
\begin{equation}
	\mathcal{E}_{\rm WS} = \mathcal{E}_{\rm e} + \mathcal{E}_{\rm g} \left(1 - \phi \right) + \frac{E_{\rm ion}}{V_{\rm WS}} \ , 
	\label{eq:ews}
\end{equation}
where $\mathcal{E}_{\rm g} = n_{\rm g} e_{\rm B} (n_{\rm g}, 1)$ ($\mathcal{E}_{\rm e}$) is the energy density of a uniform pure neutron (electron) gas at density $n_{\rm g}$ ($n_{\rm e}$), including the mass of neutrons (electrons), and the bulk interaction between the cluster and the neutron gas is treated in the excluded-volume approximation.

Following~\cite{CarreauEPJA2019}, the optimal beta-equilibrated composition of the inner crust in its ground state is determined variationally. Specifically, the WS-cell energy density in Eq.~\eqref{eq:ews} needs to be minimized under the constraint of baryon number conservation 
\begin{equation}
	n_{\rm B} = \dfrac{A}{V_{\rm WS}} + n_{\rm g}\left( 1 - \phi \right) \ ,
\end{equation}
and charge neutrality holding in every cell, see Refs.~\cite{CarreauEPJA2019,ThiAA2021,DavisAA2024} for details.

\section{Results}
\label{sec:thermo}
\subsection{Homogeneous matter: \\ low-density correction and thermodynamical properties}

We begin by examining the impact of the \textit{ab-initio}-informed low-density correction on the MM approach and its consequences for general thermodynamical quantities.
Figure~\ref{fig:e_efg_abinitio_MM_MU} shows the energy per nucleon in PNM, normalized to the Fermi gas energy $(e_{\rm B} / t_{\rm FG, n})$, as a function of $(a_{\rm n}k_{\rm F, n})$. The results are obtained using several EDF parameter sets within the MM approach (thin dashed lines), and by employing the YGLO functional for $n_{\rm B} \le n_{\rm B}^{\chi} = 0.02$~fm$^{-3}$, smoothly blended up to $n_{\rm B}^{\rm MM}$ (thick solid lines). Two cases are considered: $n_{\rm B}^{\rm MM} = 0.08$~fm$^{-3}$ (left panel) and $n_{\rm B}^{\rm MM} = 0.15$~fm$^{-3}$ (right panel).

The benchmark \textit{ab initio} results of~\cite{PalaniappanPRC2025} are shown. 
In Fig.~\ref{fig:e_efg_abinitio_MM_MU}, the standard MM predictions deviate markedly from the expected behavior in the dilute PNM regime, regardless of the chosen empirical parameter set. In contrast, we can see that our blending procedure successfully captures the correct low-density trend: the curves obtained by applying Eq.~\eqref{eq:interpolation} closely follow the AFDMC pseudodata of~\cite{GandolfiCM2022} and remain largely consistent with the MBPT3 calculations of~\cite{CoraggioPRC2013,PalaniappanPRC2023,PalaniappanPRC2025}.
Although some artificial curvature effects may appear due to the blending, they are significantly mitigated when extending the matching endpoint $n_{\rm B}^{\rm MM}$.

\begin{figure*}[tbp!]
	\includegraphics[width=0.48\linewidth]{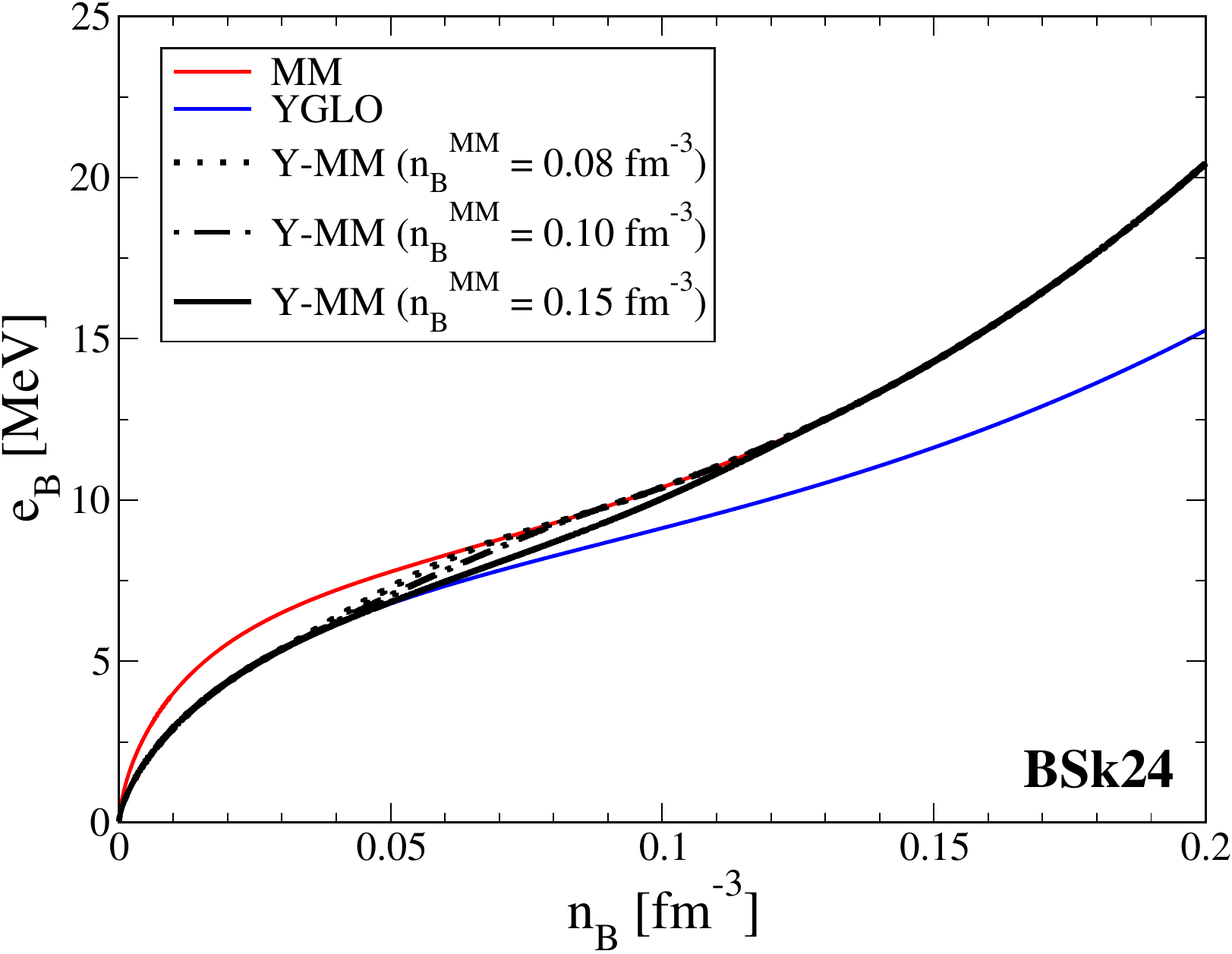} \quad \includegraphics[width=0.48\linewidth]{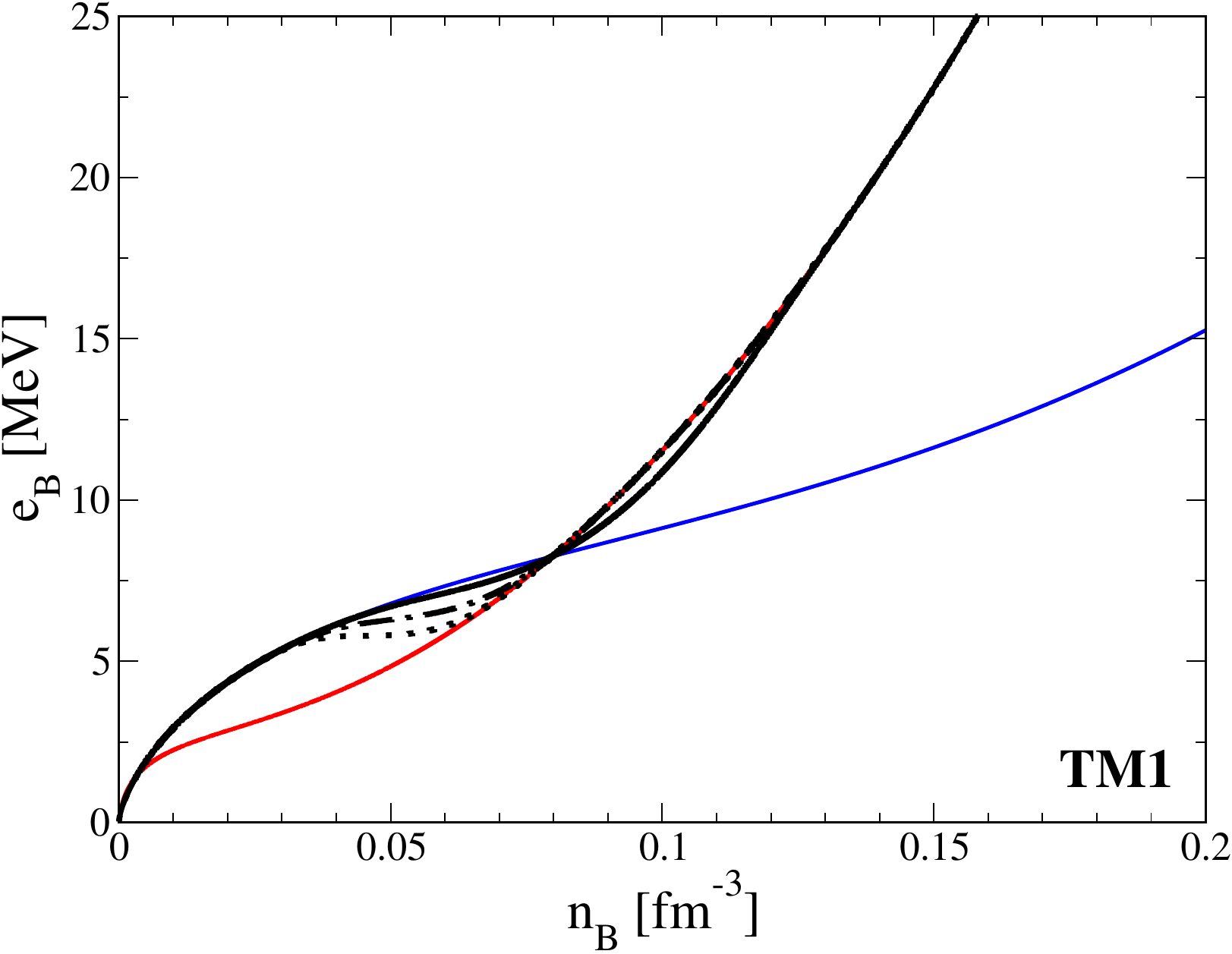}
	\caption{Energy per nucleon $e_{\rm B}$, defined in Eq.~\eqref{eq:interpolation}, as a function of the baryon number density $n_{\rm B}$ in PNM, computed within the standard MM (red lines), using the YGLO parameterization (blue lines), or with the blended functional Y-MM (black lines) for different values of $n_{\rm B}^{\rm MM}$. Left and right panels correspond to the BSk24 and TM1 empirical parameter sets, respectively.}
	\label{fig:eb_nb_PNM_MM_MU}
\end{figure*}

To get a deeper insight on the way the blending procedure works, Fig.~\ref{fig:eb_nb_PNM_MM_MU} displays the energy per nucleon in PNM obtained with the standard MM (red lines), the YGLO functional (blue lines), and the blended construction (black lines), hereafter denoted as Y-MM. The left and right panels correspond to the BSk24 and TM1 sets of empirical parameters, respectively. In both panels, we vary the endpoint of the transition region $n_{\rm B}^{\rm MM}$.
In the left panel of Fig.~\ref{fig:eb_nb_PNM_MM_MU}, the MM and YGLO curves remain close across the entire sub-saturation density range. Consequently, the Y-MM results show little sensitivity to the choice of $n_{\rm B}^{\rm MM}$, and the YGLO curve always lies below the MM BSk24 prediction, with no crossing between the two.
In contrast, the right panel of Fig.~\ref{fig:eb_nb_PNM_MM_MU} exhibits a noticeable different situation. Here, the MM and YGLO curves (red and blue, respectively) show significant discrepancies across the full range of densities, not just in the dilute regime. As a result, the Y-MM functional becomes more sensitive to the selected final endpoint. Notably, the presence of a crossing around $n_{\rm B} %= n_{\rm B}^{\rm c} 
\simeq 0.08$~fm$^{-3}$ leads to a mismatch in the slopes of the two curves, implying differences in the density derivatives of the energy.

Therefore, the blending may introduce changes in curvature, which could induce unphysical features in the second-order density derivatives of the energy per nucleon, leading to spurious instabilities.

\begin{figure}[tbp!]
	\includegraphics[width=\linewidth]{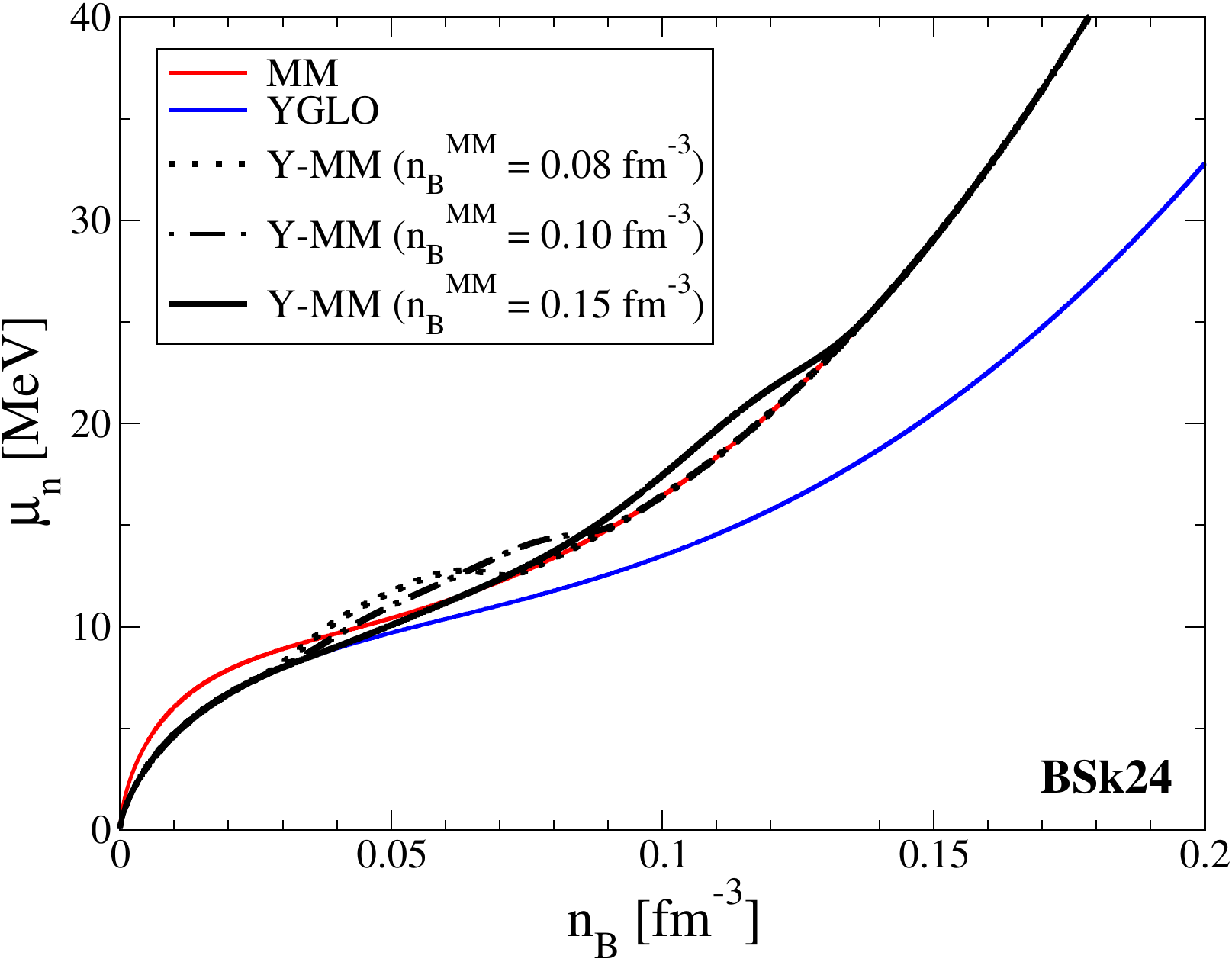}
	\caption{Neutron chemical potential $\mu_{\rm n}$ as a function of the baryon number density $n_{\rm B}$ in PNM, obtained within the MM approach (red lines), using the YGLO functional (blue lines), or with the blended Y-MM functional (black lines), for different values of $n_{\rm B}^{\rm MM}$. Results are shown for the BSk24 empirical parameter set.}
	\label{fig:mu}
\end{figure}

This effect is more clearly seen in the density dependence of the neutron chemical potential $\mu_{\rm n}$, defined in Eq.~\eqref{eq:muq}, and shown in Fig.~\ref{fig:mu}. Results from the MM approach (red line) are compared with those from the standard YGLO functional (blue line) and from the blended Y-MM construction (black lines) for different values of the endpoint $n_{\rm B}^{\rm MM}$. For the representative case of the BSk24 parameter set, one observes that if the blending region is too narrow (i.e., $n_{\rm B}^{\rm MM} \lesssim 0.08$~fm$^{-3}$), the neutron chemical potential, although continuous, develops a segment with negative slope.

Such spurious instabilities can arise from the blending procedure defined in Eq.~\eqref{eq:interpolation}, even when the energy densities of the MM and YGLO models do not cross. This behavior is generic and occurs for any set of empirical MM parameters if the blending range $ [n_{\rm B}^{\chi} , n_{\rm B}^{\rm MM} ]$ is too limited. Nevertheless, these unphysical oscillations are significantly reduced when the transition region is extended by increasing~$n_{\rm B}^{\rm MM}$.

\begin{figure}[tbp!]
	\includegraphics[width=\linewidth]{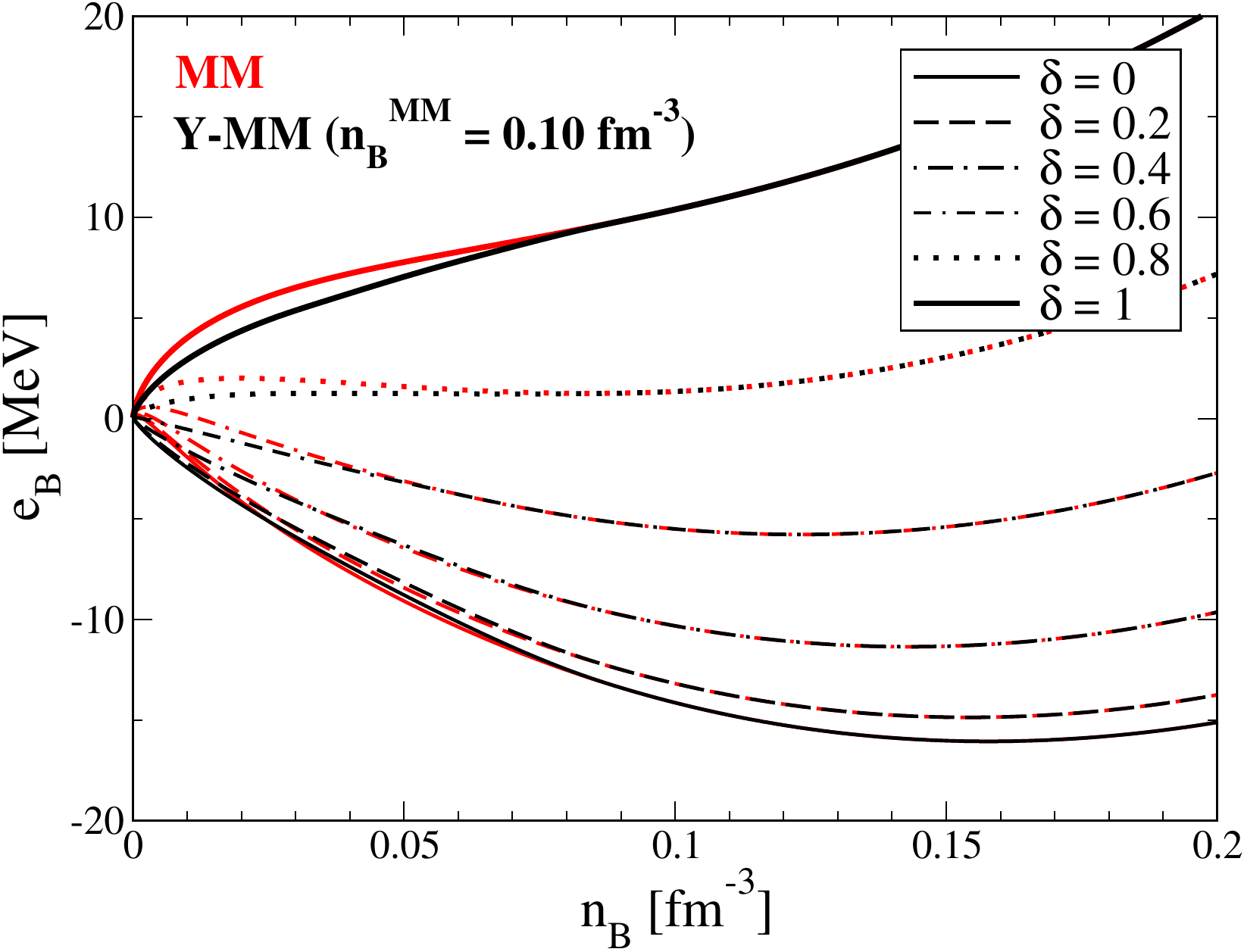}
	\caption{Energy per nucleon $e_{\rm B}$ as a function of the total baryon number density $n_{\rm B}$ for different values of the isospin asymmetry $\delta$,  as obtained by the standard MM (red lines) or with the functional Y-MM, with $n_{\rm B}^{\rm MM} = 0.10$ fm$^{-3}$ (black lines). Results are shown for the BSk24 empirical parameter set.}
	\label{fig:eb_nb_delta_MM_MU}
\end{figure}

Furthermore, the blending strategy defined in Eq.~\eqref{eq:interpolation} is applied for arbitrary isospin asymmetry. Figure~\ref{fig:eb_nb_delta_MM_MU} displays the energy per baryon as a function of the total baryon number density for various values of the asymmetry $\delta$, computed using the blended Y-MM functional (black lines). The results correspond to the BSk24 empirical parameter set, with the final endpoint fixed at $n_{\rm B}^{\rm MM} = 0.10$~fm$^{-3}$. For comparison, the standard MM predictions (red lines) are also shown.

The most pronounced differences between the two models appear in the PNM limit ($\delta = 1$), as expected. However, deviations are also visible at all intermediate asymmetries, including the SNM case ($\delta = 0$). As a consequence, the blended construction affects general thermodynamical properties across a wide range of asymmetries, and may thus influence the predicted composition of the inner crust. In particular, the region where Y-MM and MM results diverge includes densities and asymmetries characteristic of the neutron gas in the inner crust~\cite{burPRC2015}, extending up to the CC transition.

\subsection{Inhomogeneous matter: \\ isotopic composition of the inner crust}

\begin{figure*}[tbp!]
	\includegraphics[width=.8\linewidth]{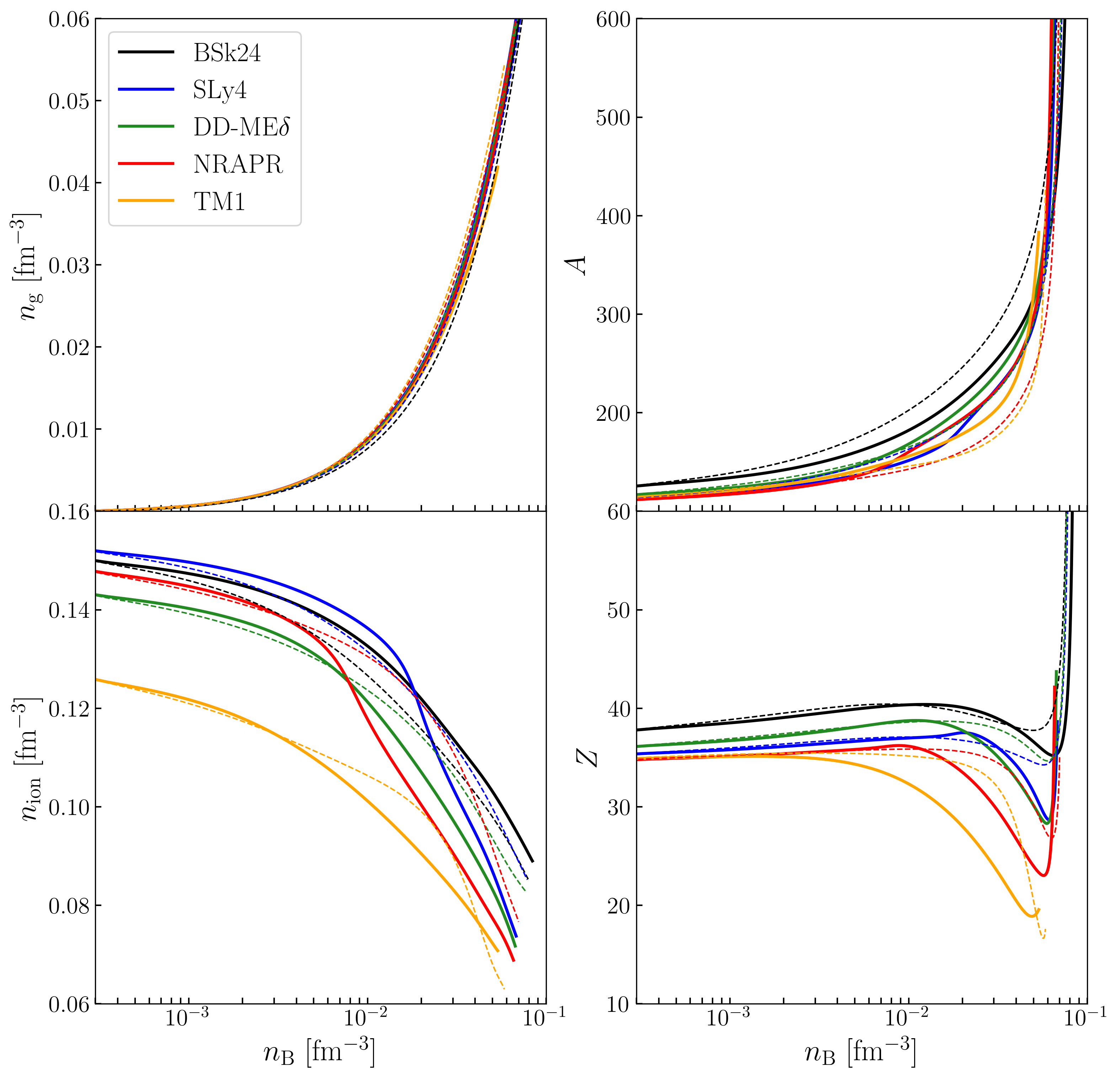} 
	\caption{Free neutron-gas density $n_{\rm g}$ (top left panel), ion density (bottom left panel), cluster mass $A$ (top right panel), and proton number $Z$ (bottom right panel) numbers, in the inner crust as a function of $n_{\rm B}$ for different set of empirical parameters of EDFs available in literature. Results are obtained within the standard MM (thin dashed lines), or with the functionals Y-MM (thick full lines). 
	}
	\label{fig:composition_MU}
\end{figure*}

As previously noted, even when uncertainties from surface parameters are neglected, the crust composition remains highly model dependent due to differences in the bulk matter description. Figure~\ref{fig:composition_MU} shows the predictions from the standard MM (dashed thin lines) for the free neutron-gas density $n_{\rm g}$, the ion density $n_{\rm ion}$, cluster mass number $A$, and proton number $Z$ in the inner crust, using various empirical parameter sets from widely used EDFs. As density increases, more neutrons drip into the surrounding gas, leading to a rise in $n_{\rm g}$. At the same time, the cluster mass number increases up to $A \sim 600$, while the proton number remains approximately constant at $Z \sim 40$, in agreement with previous works~\cite{Pearson2018,CarreauAA2020,MondalPRC2020}. These trends persist up to the CC transition, beyond which the cluster and gas densities merge and the curves terminate. Notably, the MM approach exhibits a significant spread in cluster sizes, reflecting the bulk EoS uncertainties.

Figure~\ref{fig:composition_MU} also shows the results obtained using the Y-MM functionals for the energy per baryon, as defined in Eq.~\eqref{eq:interpolation}. The low-density regime is found to play a crucial role in shaping the crustal composition. In particular, adopting the Y-MM prescription significantly reduces the spread in neutron-gas density and cluster size associated with different empirical EDF parameter sets, across the full density range relevant to the inner crust.

As seen in Fig.~\ref{fig:e_efg_abinitio_MM_MU}, and reflected in the top panel of Fig.~\ref{fig:composition_MU}, the neutron gas is energetically more favored under the Y-MM functional than with the standard MM for BSk24, while the opposite holds for EDFs such as NRAPR or TM1. This behavior reduces the spread in cluster size across the entire inner crust, up to the CC transition. Specifically, when more (fewer) neutrons populate the gas in the case of BSk24 (TM1), fewer (more) neutrons remain in the clusters, leading to a corresponding decrease (increase) of~$A$.

Concerning the behavior of the proton number $Z$ and the cluster density $n_{\rm ion}$, for all models the Y-MM curves predict a more pronounced dip with increasing density. The correlation between $Z$ and $n_{\rm ion}$ is imposed by the Baym virial theorem
$E_{\rm surf}=2 E_{\rm Coul}$ implicit in the variational equations \cite{CarreauEPJA2019}.
In turn, the systematic effect on $n_{\rm ion}$ can be understood from the effect of the YGLO correction. As it can be appreciated from Fig.~\ref{fig:e_efg_abinitio_MM_MU}, the correction tends to reduce the energy difference between the
dilute and dense phase, and therefore their density difference as function of $n_{\rm B}$. 

These findings are in good qualitative agreement with the results of~\cite{GuptaPRC2024}. In that work, the \textit{ab initio} functional of~\cite{PalaniappanPRC2023} was adopted for the neutron-gas component, while two different Skyrme forces were used for the cluster. A significant difference in the crust composition and EoS was observed when the cluster functional deviated strongly from the \textit{ab initio} behavior. In contrast, this difference was reduced when a functional consistent with the \textit{ab initio} predictions above $n_{\rm B} \gtrsim 0.02$~fm$^{-3}$ was used. A similar trend emerges in our unified treatment when comparing the results for NRAPR (or TM1) and DDME$\delta$.

\section{Bayesian analysis}
\label{sec:bayesian}

The Bayesian framework allows to update prior beliefs on a given quantity, with the constraints arising from multiple sources, on the basis of the formula:
\begin{equation}
	p_{\rm post}(\mathbf {X}|\textrm{data}) \propto p(\textrm{data}|\mathbf{X}) p_{\rm prior}(\mathbf{X}), 
	\label{eq:bayes}
\end{equation}
where $\mathbf{X}$ denotes the set of parameters of our (Y-)MM approach, $p_{\rm prior}(\mathbf{X})$ is the prior probability density function (PDF) of $\mathbf{X}$, $p(\textrm{data}|\mathbf{X})$ the likelihood of observing the data for the same parameter set, and $p_{\rm post}(\mathbf{X}|\textrm{data})$ the (unnormalized)  posterior PDF. Equation~\eqref{eq:bayes} is used to quantify the uncertainty on the crust properties induced by our imperfect knowledge of the baryonic bulk energy in the dilute regime. 

\subsection{Informed prior and posterior distribution}

\begin{table}[btp!]
	\centering
	\renewcommand{\arraystretch}{1.2}
	\begin{tabular}{c | c c}
		\toprule
		$X_{k}$                      & $X_{k}^{\rm min}$ & $X_{k}^{\rm max}$ \\
		\midrule
		$n_{\text{sat}}$ [fm$^{-3}$] & 0.15              & 0.17              \\
		$E_{\text{sat}}$ [MeV]       & -17               & -15               \\
		$K_{\text{sat}}$ [MeV]       & 190               & 270               \\
		$Q_{\text{sat}}$ [MeV]       & -1000             & 1000              \\
		$Z_{\text{sat}}$ [MeV]       & -3000             & 3000              \\
		$E_{\text{sym}}$ [MeV]       & 26                & 38                \\
		$L_{\text{sym}}$ [MeV]       & 10                & 80                \\
		$K_{\text{sym}}$ [MeV]       & -400              & 200               \\
		$Q_{\text{sym}}$ [MeV]       & -2000             & 2000              \\
		$Z_{\text{sym}}$ [MeV]       & -5000             & 5000              \\
		$m^*_{\text{sat}}/m$         & 0.6               & 0.8               \\
		$\Delta m^*_{\text{sat}}/m$  & 0.0               & 0.2               \\
		\bottomrule
	\end{tabular}
	\caption{Minimum (maximum) values $X_{k}^{\rm min}$ ($X_{k}^{\rm max}$) of each parameter $X_{k}$, with $k = 1, \dots, 2 (\mathcal{N} + 2)$ ($\mathcal{N} = 4$) of the parameter set $\textbf{X}$, in the prior distribution.}
	\label{tab:prior}
\end{table}

We first generate a large set of EoSs defined by parameters $\mathbf{X}$. This constitutes a sample of a prior, which is filtered through physical and astrophysical constraints to build an \textit{informed prior} (IP), which constitutes the baseline for all analyses in this work. The resulting posterior distribution may further incorporate additional filters, as detailed below. The dimensionality of our parameter space is \(2(\mathcal{N} + 2)\), with \(\mathcal{N} = 4\).

The flat prior of \(\mathbf{X}\) is defined as a uniform distribution for each parameter within physically reasonable bounds~\cite{ThiUni2021,MontefuscoAA2025}, 
\begin{equation}
	p_{\text{prior}}(\textbf{X}) = \prod_{k=1}^{2(\mathcal{N}+2)} f (X^{\text{min}}_k, X^{\text{max}}_k; X_k),
\end{equation}
where \(f\) is uniform between \(X^{\text{min}}_k\) and \(X^{\text{max}}_k\) (see Table~\ref{tab:prior}).

The \emph{informed prior} (IP) is defined as:
\begin{equation}
	p_{\text{IP}}(\textbf{X}) \propto w_{\rm phys}(\textbf{X})\, w_{\rm Mmax}(\textbf{X})\, e^{-\chi^2(\textbf{X})/2} \, p_{\text{prior}}(\textbf{X}),
\end{equation}
where: (i) \(w_{\rm phys}(\textbf{X})\) is a sharp filter that enforces general physical conditions (applied on the functional defined in Eq.~\eqref{eq:interpolation}) such as a positive symmetry energy, thermodynamic stability, and a subluminal barotropic sound speed (see~\cite{MontefuscoAA2025} for a detailed discussion);
(ii) \(w_{\rm Mmax}(\textbf{X})\) ensures that the EoS supports a maximum mass $M_{\text{max}} \gtrsim 1.97\, M_{\odot}$ ($M_\odot$ being the solar mass)~\cite{AntoniadisSCI2013};
(iii) \(\chi^2(\textbf{X})\) corresponds to the fit of nuclear masses from the Atomic Mass Evaluation (AME)~\cite{WangCPC2012}, used to determine the surface parameters \((\sigma_0, \sigma_{0,c}, b_s, \beta)\) for each EoS, following~\cite{CarreauEPJA2019,MontefuscoAA2025}. The choice to consider the maximum mass as the only astrophysical filter is supported by earlier studies indicating that additional constraints from NICER and LIGO-Virgo-KAGRA have a negligible impact on crustal properties~\cite{ThiEPJA2021,klausner2025arXiv}.

To isolate the impact of low-density constraints, we optionally apply an additional filter based on chiral EFT, thus defining the posterior distribution:
\begin{equation}
	p_{\text{post}}(\textbf{X}) \propto w_{\rm EFT}(\textbf{X}) \, p_{\text{IP}}(\textbf{X}),
\end{equation}
where \(w_{\rm EFT}(\textbf{X})\) requires the PNM energy per particle in the density range \(n_{\rm B} \in [0.02, 0.20]\,\text{fm}^{-3}\) to lie within a 5\%-enlarged band around the N2LO \textit{ab initio} predictions of~\cite{HuthNAT2022}~(see their Fig.~4), which include local 2N and 3N EFT interactions~\cite{LynnPRL2016}.
This is the only constraint that is toggled in the present study.
No additional filters are imposed on the SNM behavior, as the uncertainty in the energy per particle and pressure near saturation within the prior is already comparable to, or narrower than, that of \textit{ab initio} predictions~\cite{CarreauEPJA2019}. This results from tight empirical constraints derived from low-energy nuclear data, which guided the choice of parameter ranges in Table~\ref{tab:prior}. In contrast, the prior leaves the PNM EoS largely unconstrained, yielding broader uncertainties than those from \textit{ab initio} calculations.

% \begin{table}[tbp!]
% 	\centering
% 	\begin{tabular}{c|c}
% 		\toprule
% 		\textbf{Model} %& IP [\%] & Post [\%] 
% 		                      & Post/IP [\%] \\
% 		\midrule
% 		MM                    & %11.18 & 0.31           
% 		                      & 2.78         \\
% 		\midrule
% 		Y-MM (0.08 fm$^{-3}$) & %0.44 & 0.16 &           
% 		36.36 \\
% 		Y-MM (0.10 fm$^{-3}$) & %0.96 & 0.36 &           
% 		37.50 \\
% 		Y-MM (0.12 fm$^{-3}$) & %1.92 & 0.74 &           
% 		38.54 \\
% 		Y-MM (0.14 fm$^{-3}$) & %3.58 & 1.33 &           
% 		37.15 \\
% 		Y-MM ($n_{\rm sat}$)  & %6.57 & 2.38 &           
% 		36.23 \\
% 		\bottomrule
% 	\end{tabular}
% 	\caption{Percentage of sets belonging to the \textit{informed prior} (IP) fulfilling the chiral constraint (Post), for the different models and different choices of the final endpoint of the interpolation $n_{\rm B}^{\rm MM}$ with YGLO considered in this work.}
% 	\label{tab:percentage}
% \end{table}

\begin{table}[tbp!]
	\centering
	\begin{tabular}{c|cc}
		\toprule
		\textbf{Model}        & Post [\%] & Post/IP [\%] \\
		\midrule
		MM                    & 0.31      & 2.78         \\
		\midrule
		Y-MM (0.08 fm$^{-3}$) & 0.16      & 36.36        \\
		Y-MM (0.10 fm$^{-3}$) & 0.36      & 37.50        \\
		Y-MM (0.12 fm$^{-3}$) & 0.74      & 38.54        \\
		Y-MM (0.14 fm$^{-3}$) & 1.33      & 37.15        \\
		Y-MM ($n_{\rm sat}$)  & 2.38      & 36.23        \\
		\bottomrule
	\end{tabular}
	\caption{Percentage of sets belonging to the \textit{informed prior} (IP) fulfilling the chiral constraint (Post) and corresponding ratio (Post/IP), for the different models and different choices of $n_{\rm B}^{\rm MM}$ within Y-MM considered in this work.}
	\label{tab:percentage}
\end{table}

This is illustrated in Table~\ref{tab:percentage}, which reports the percentage of parameter sets from the IP prior that satisfy the $\chi$-EFT filter and contribute to the posterior distribution. Both versions of the MM approach considered in this work are shown. In the \textit{ab-initio}-benchmarked version, where blending with YGLO is applied, different values of $n_{\rm B}^{\rm MM}$ are explored, including, as an extreme case, $n_{\rm B}^{\rm MM} = n_{\rm sat}$, where $n_{\rm sat}$ is the SNM saturation density predicted by each parameter set and around which the MM expansion is defined. 

To ensure consistent statistics across configurations, $6 \times 10^6$ models are sampled from the prior, yielding at least $10^4$ accepted models in each case. It is found that the $\chi$-EFT filter is especially restrictive in the standard MM scenario, with only 2.78\% of the models contributing to the IP distribution also passing the filter.

\begin{figure}[tbp!]
	\subfigure[MM (IP)]{
		\includegraphics[width=0.45\linewidth]{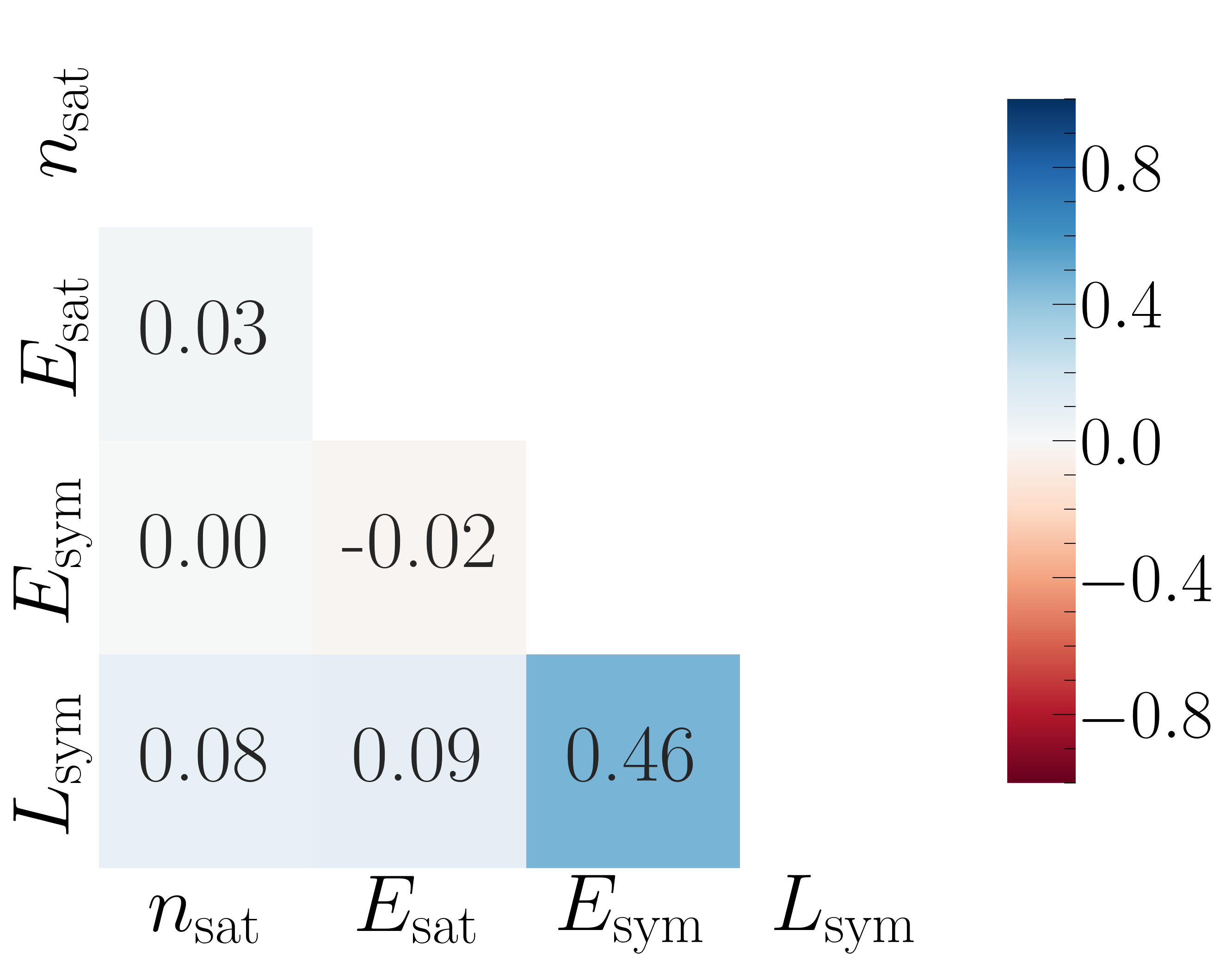}
		\label{fig:correlation_MM_nochi}
	}
	\hfill
	\subfigure[Y-MM (IP)]{
		\includegraphics[width=0.45\linewidth]{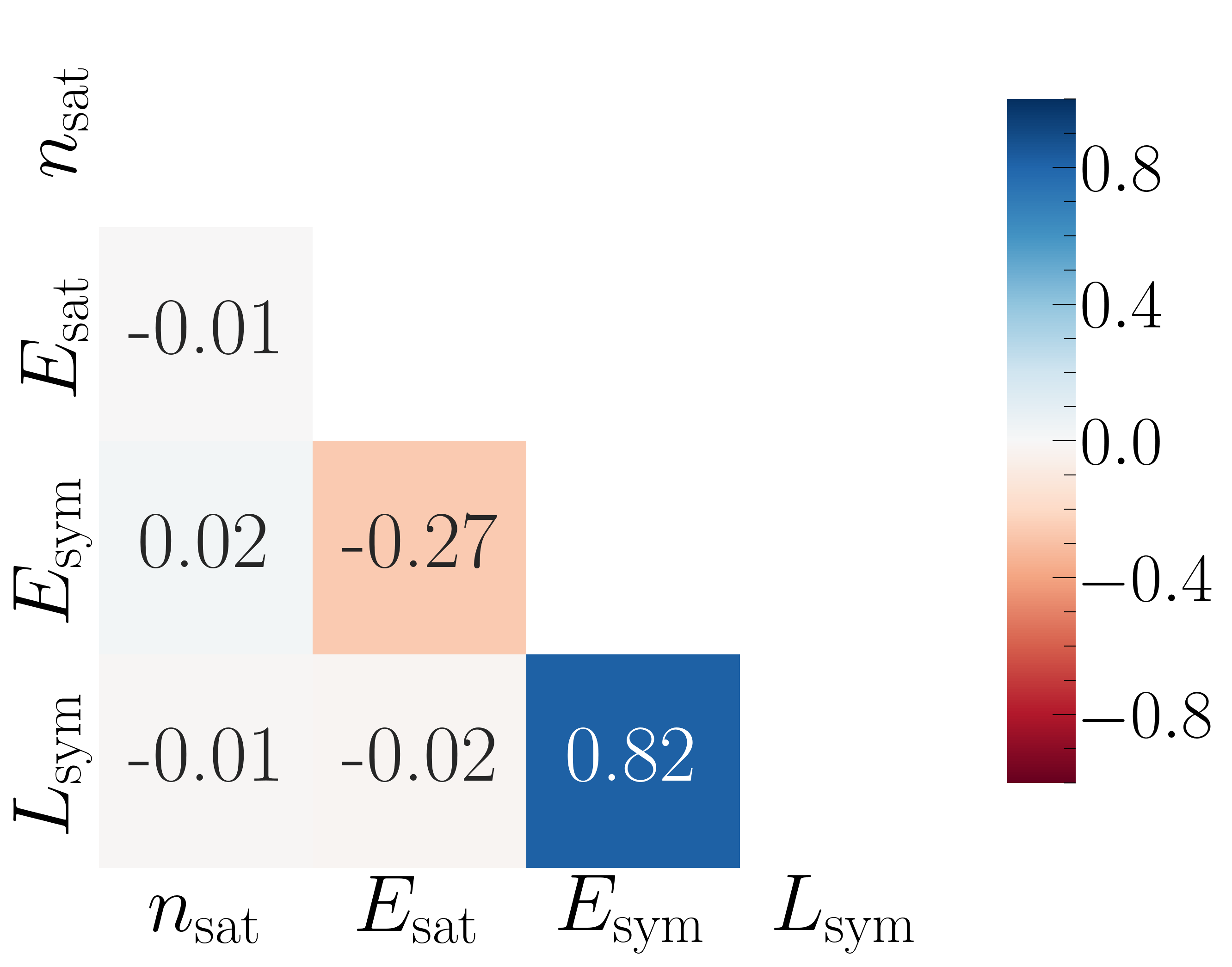}
		\label{fig:correlation_MU_nochi}
	}
	\hfill
	\subfigure[MM (Post)]{
		\includegraphics[width=0.45\linewidth]{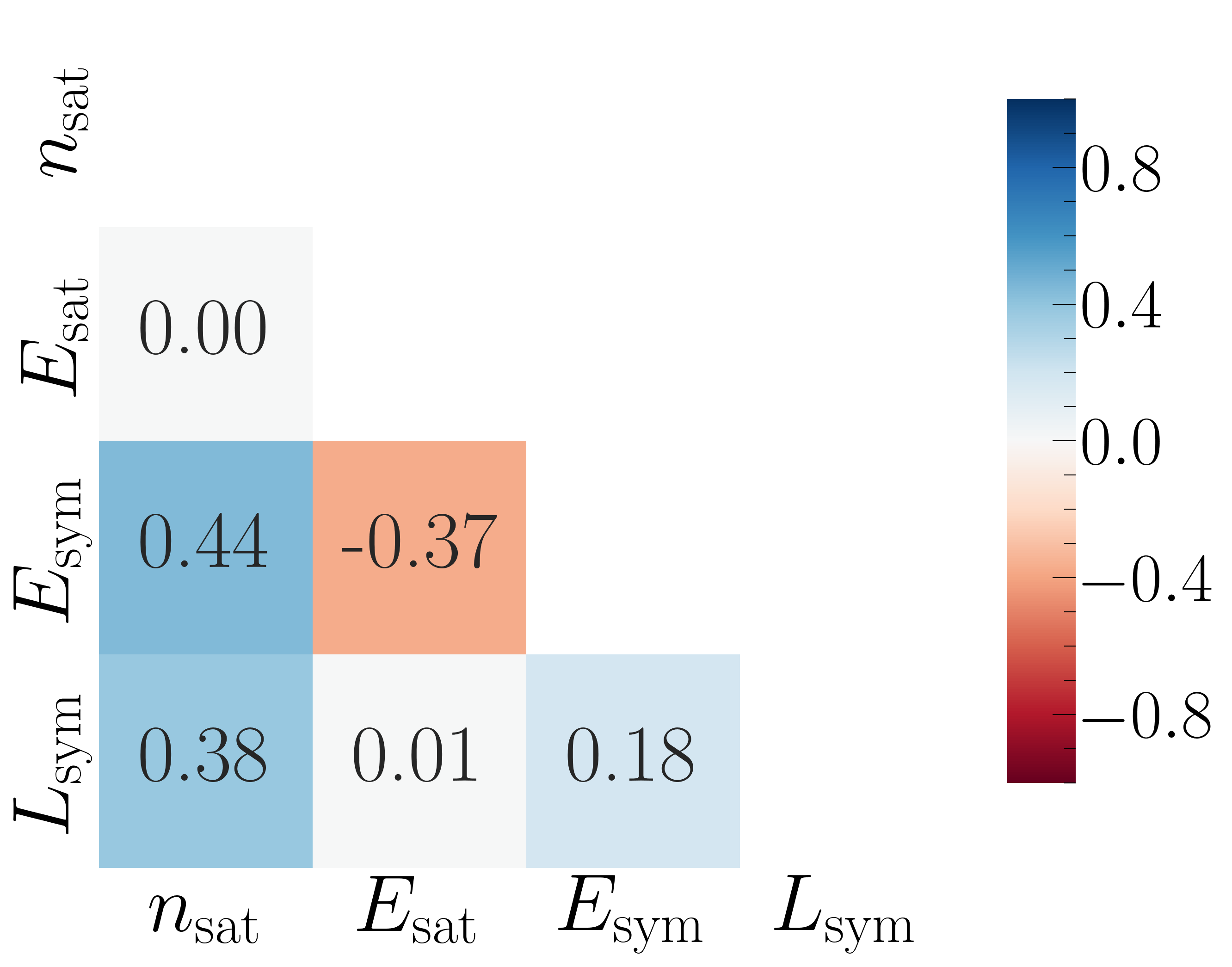}
		\label{fig:correlation_MM_chi}
	}
	\hfill
	\subfigure[Y-MM (Post)]{
		\includegraphics[width=0.45\linewidth]{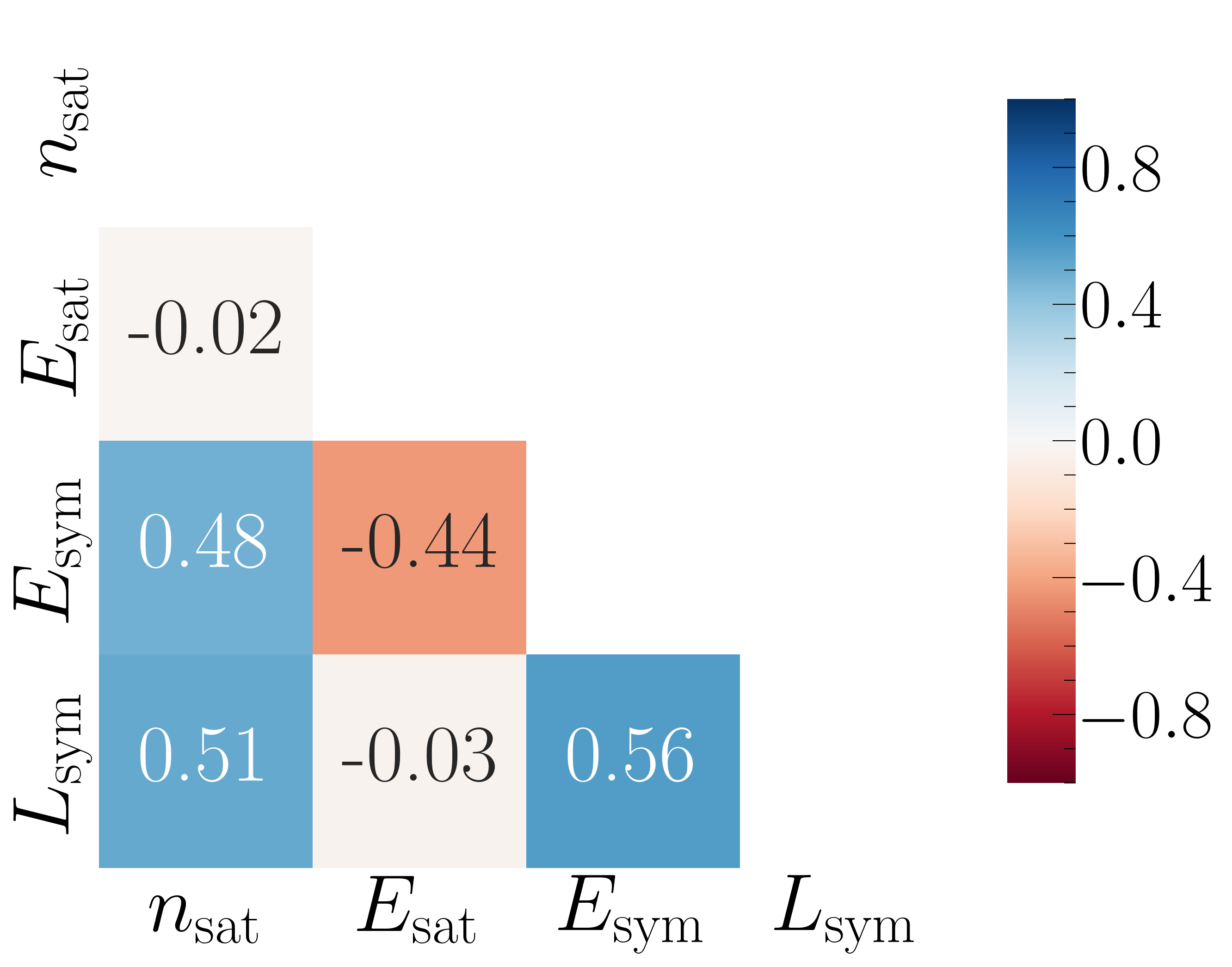}
		\label{fig:correlation_MU_chi}
	}
	\caption{Correlation matrices  as obtained in the IP (top panels) and posterior (bottom panels) distribution of the standard MM (left panels) and its \textit{ab-initio}-benchmarked version Y-MM (right panels).} 
	\label{fig:correlation_matrices}
\end{figure}
 
The $\chi$-EFT filter is significantly less restrictive in the Y-MM approach, owing to its improved treatment of low-density matter. Interestingly, the ratio between the number of models in the posterior and in the IP distribution remains approximately constant in the Y-MM case as $n_{\rm B}^{\rm MM}$ increases up to saturation density.
However, the total percentage of models contributing to the IP distribution in Y-MM may be lower than in the standard MM case. This occurs when the mechanical stability condition in PNM, given by Eq.~\eqref{eq:inequality}, is not fulfilled unless $n_{\rm B}^{\rm MM}$ is sufficiently large.

\begin{figure*}[tbp!]
	\centering
	\includegraphics[width=.7\linewidth]{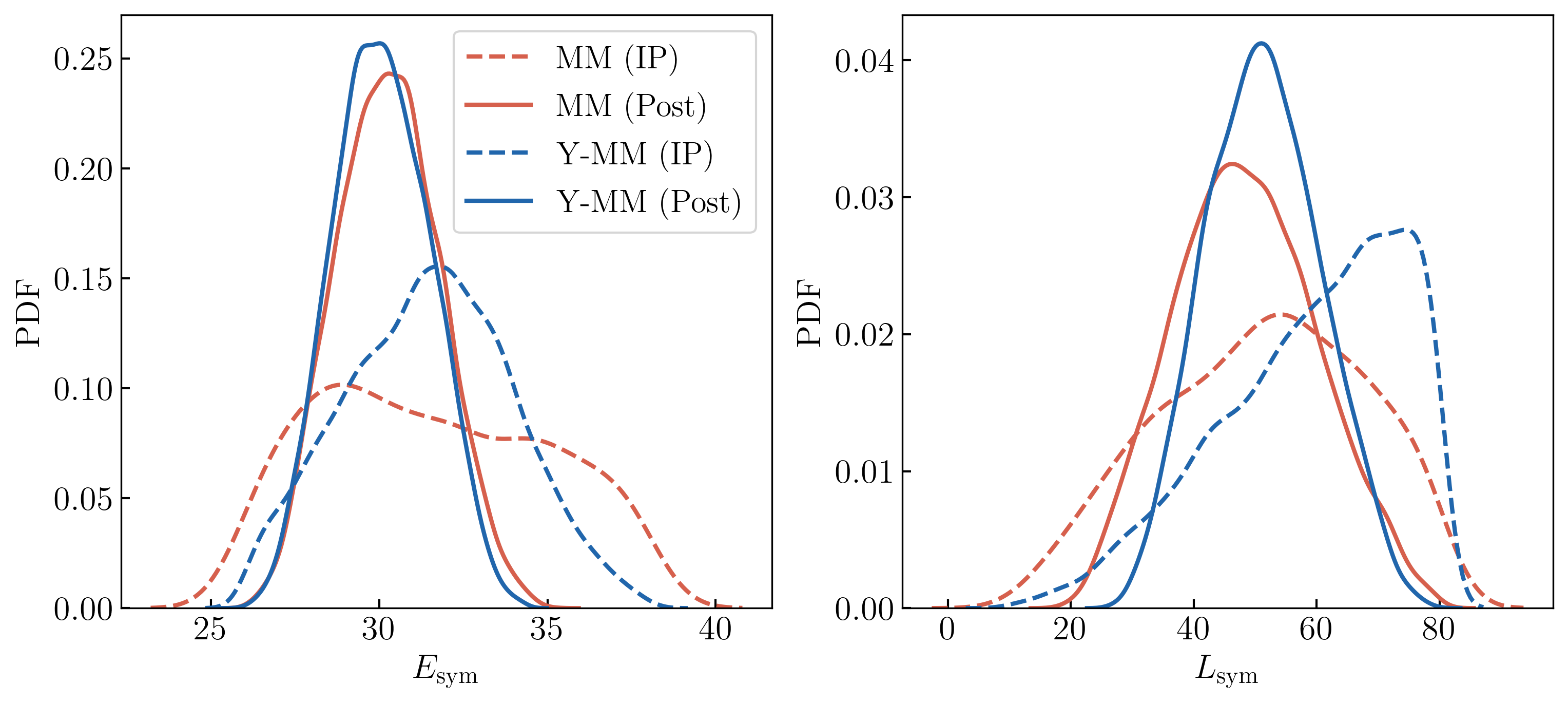}
	\caption{Probability density functions (PDFs) of the zeroth (left panel) and first (right panel) order empirical parameters in the isovector channel, extracted from the IP (dashed lines) and posterior (solid lines) distributions for the standard MM (red lines) and its \textit{ab-initio}-benchmarked version Y-MM (blue lines).}
	\label{fig:pdf}
\end{figure*}

Finally, for $n_{\rm B}^{\rm MM} \gtrsim 0.10$~fm$^{-3}$, the fraction of models in the posterior exceeds that obtained in the standard MM. This indicates that many parameter sets, which are compatible with high-density (IP) constraints, can also satisfy the tight $\chi$-EFT bounds on the PNM EoS, provided a suitable correction is implemented in the subsaturation regime. These sets were excluded in previous analyses, missing the opportunity to include them in Bayesian posteriors.

\subsection{Correlations analysis of set parameters}

Before turning to predictions for the NS EoS and related observables, we examine the correlations embedded within the (Y-)MM approaches, both at the level of the IP-filtered distributions and in the full posterior PDFs. 
Given a pair of parameters $(X_i, X_j)$, the Pearson correlation coefficient is defined as
\begin{equation}
	r_{ij} = \frac{\text{cov}(X_i, X_j)}{\sigma_{X_i} \, \sigma_{X_j}} \,,
\end{equation}
where $\text{cov}(X_i, X_j)$ is the covariance and $\sigma_{X_i}$ ($\sigma_{X_j}$) is the standard deviation of the parameter $X_i$ ($X_j$).

Figure~\ref{fig:correlation_matrices} shows the correlation matrices for the empirical parameters associated with the zeroth and first orders of the MM expansion. As noted in earlier applications of the MM, the IP filters induce nontrivial correlations among the isovector parameters, particularly between $E_{\text{sym}}$ and $L_{\text{sym}}$, a well-documented trend in the literature~\cite{LimEPJA2019,KlausnerPRC2025}. 
This correlation is further reinforced in the Y-MM (IP) case, due to the matching with YGLO. The chiral filter introduces an anticorrelation between $E_{\text{sat}}$ and $E_{\text{sym}}$, along with a positive correlation between $E_{\text{sym}}$ (or $L_{\text{sym}}$) and $n_{\text{sat}}$, which are both moderately enhanced in the Y-MM posterior.

These correlations are reflected in the PDFs of the zeroth- and first-order isovector parameters, shown in Fig.~\ref{fig:pdf}. The distributions correspond to the IP (dashed lines) and posterior (solid lines) PDFs for the standard MM (red) and the \textit{ab initio}-benchmarked Y-MM (blue), respectively. In the standard MM, the physical and astrophysical constraints implemented via $w_{\rm IP}$ exert minimal influence on the low-order empirical parameters, consistent with previous findings~\cite{MargueronPRC2018,CarreauEPJA2019}. In contrast, the IP PDFs in the Y-MM case become more peaked and shift toward larger central values, with the most pronounced effect seen for~$L_{\rm sym}$.

Unlike the IP constraints, the chiral filter plays a central role in constraining $E_{\rm sym}$ and $L_{\rm sym}$. The widths of the PDFs further decrease in the Y-MM posterior, indicating stronger parameter determination. As a result, the posterior mean value for $E_{\rm sym}$ in the Y-MM approach closely matches that of the standard MM (see Table~\ref{tab:pdf_stats}), while the central value of $L_{\rm sym}$ shifts moderately. This shift arises because the Y-MM functionals lie close to the lower boundary of the chiral band, and a stiffer symmetry energy around saturation is needed to satisfy the IP constraints. Additionally, the PDF dispersion is reduced, especially for~$L_{\rm sym}$.

\begin{table}[tbp!]
	\centering
	\begin{tabular}{c|c|c}
		\toprule
		\bf{Model}  & $E_{\rm{sym}}$ [MeV] & $L_{\rm{sym}}$ [MeV] \\
		\midrule
		MM (IP)     & 31.59 $\pm$ 3.38     & 51.07 $\pm$ 16.45    \\
		Y-MM (IP)   & 31.47 $\pm$ 2.47     & 59.25 $\pm$ 14.96    \\
		MM (Post)   & 30.27 $\pm$ 1.52     & 48.43 $\pm$ 11.53    \\
		Y-MM (Post) & 30.09 $\pm$  1.43    & 51.36 $\pm$ 9.21     \\
		\bottomrule
	\end{tabular}
	\caption{Mean value and standard deviation of the PDFs of the zeroth- and first-order empirical parameters in the isovector channel, as obtained in the IP and posterior distributions of the standard MM and its \textit{ab-initio}-benchmarked version Y-MM.}
	\label{tab:pdf_stats}
\end{table}

\begin{figure*} [tbp!]
	\includegraphics[width=.45\linewidth]{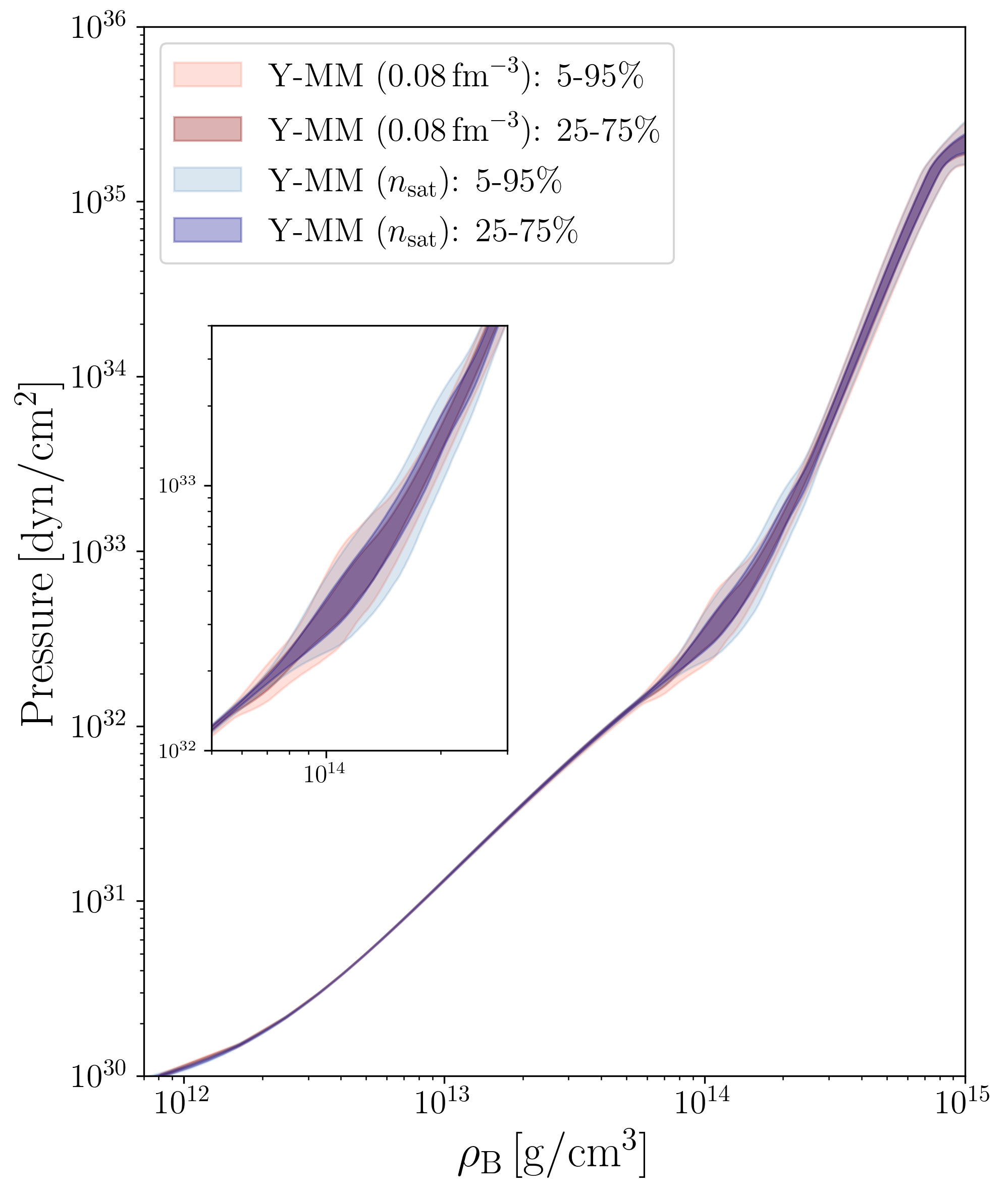} \hfill
	\includegraphics[width=.45\linewidth]{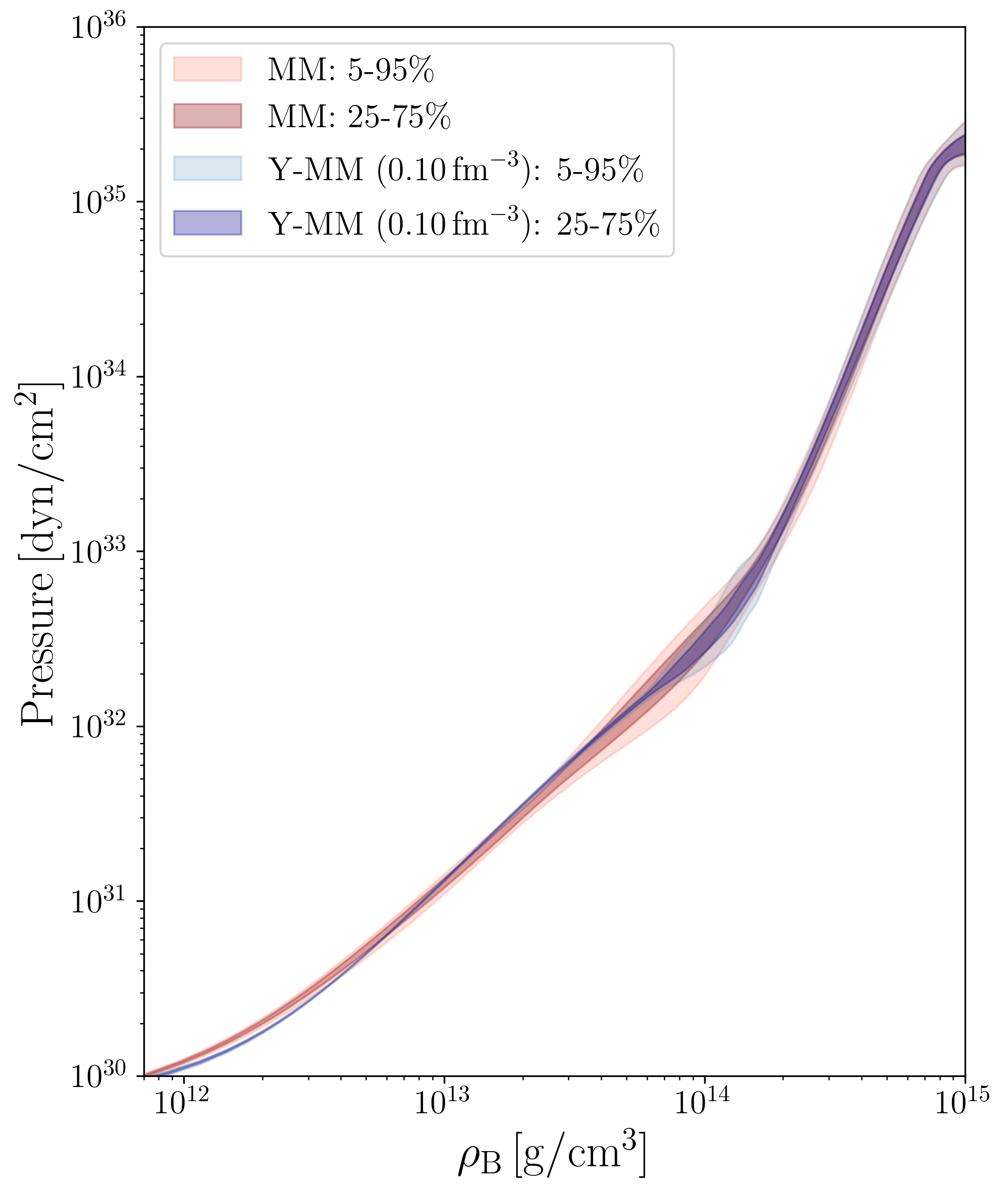}  
    \caption{Left panel: Pressure as a function of the total baryon mass density $\rho_{\rm B}$, obtained via Bayesian inference within the Y-MM approach, for the two extreme values of $n_{\rm B}^{\rm MM}$ considered in this work: $0.08~\mathrm{fm}^{-3}$ (red bands) and $n_{\rm sat}$ (blue bands). Right panel: Comparison between the standard MM (red bands) with its \textit{ab-initio}-benchmarked counterpart, Y-MM (blue bands), for a fixed value $n_{\rm B}^{\rm MM} = 0.10~\mathrm{fm}^{-3}$. }
	\label{fig:pressure}
\end{figure*}

\subsection{Neutron-star Equation of State}

We now examine the impact of our refined treatment of the PNM EoS at subsaturation densities on general astrophysical properties, starting with the NS EoS.
Figure~\ref{fig:pressure} shows the pressure as a function of the total baryon mass density, $\rho_{\rm B} = m n_{\rm B}$, as inferred from the posterior distribution. In the left panel, color bands correspond to different choices of $n_{\rm B}^{\rm MM}$ within the Y-MM approach. For each $n_{\rm B}^{\rm MM}$ value, two band intensities are displayed, indicating the confidence percentiles specified in the legend. The NS EoS shows little sensitivity to $n_{\rm B}^{\rm MM}$: the bands are extremely narrow throughout the inner crust, at least up to $\rho_{\rm B} \simeq 5 \times 10^{13}$~g/cm$^3$, and begin to widen only near the CC transition. Some sensitivity appears in that region, as shown in the inset, which zooms in on the CC transition point.

In particular, the blue bands (larger $n_{\rm B}^{\rm MM}$) remain narrower across a wider range of the inner crust than the red bands, reflecting the fact that the Y-MM functional for $n_{\rm B}^{\rm MM} = 0.08$~fm$^{-3}$ departs from YGLO at lower densities. Conversely, near saturation density ($\rho_{\rm B} \simeq 2 \times 10^{14}$~g/cm$^3$), the blue bands are broader, due to the higher fraction of retained models in the posterior for this $n_{\rm B}^{\rm MM}$ choice. Based on this analysis, we fix $n_{\rm B}^{\rm MM} = 0.10$~fm$^{-3}$ in the following, allowing us to isolate the effect of benchmarking against microscopic calculations at low density while keeping the number of models in the posterior comparable between the MM and Y-MM cases (see Table~\ref{tab:percentage}).

The right panel of Fig.~\ref{fig:pressure} compares the NS EoS obtained from the posterior distributions of the standard MM (red bands) and its \textit{ab-initio}-benchmarked counterpart, Y-MM (blue bands), with $n_{\rm B}^{\rm MM} = 0.10$~fm$^{-3}$. The most prominent feature is the narrowing of the blue bands across the entire inner crust region, confirming the reduction of model dependence discussed in Sec.~\ref{sec:inhomogeneous}.

In addition to this overall reduction, a distinct trend emerges in the low-density outer layers of the inner crust, where the blue bands diverge from the red ones. This deviation reflects the difference between the heuristic extrapolation to zero density in the standard MM and the \textit{ab initio} benchmark implemented in Y-MM. At suprasaturation densities, however, both approaches converge, as the high-density behavior is governed by the IP constraints in all cases. 

\subsection{Inference of crustal properties: CC density transition}

We now present the results for the pressure $P_{\rm CC}$ at the CC transition density $n_{\rm CC}$, analyzing the behavior of the (Y-)MM approach and the role played by the chiral filter. Following the methodology adopted in Refs.~\cite{CarreauPRC2019,ThiUni2021}, we define the CC transition point ``from the crust''~\cite{CarreauEPJA2019}, identifying $n_{\rm CC}$ as the density at which the energy density of the inner crust (within the WS cell, Eq.~\eqref{eq:ews}) equals that of homogeneous matter in the core, under conditions of beta equilibrium.

The robustness of the inferred predictions with respect to prior choices has been addressed in earlier MM studies of the CC transition~\cite{CarreauEPJA2019,ThiUni2021}, showing negligible sensitivity to the choice of prior, once physical and astrophysical constraints are enforced.

\begin{figure}[tbp!]
	\centering
	\subfigure[MM (IP)]{
		\includegraphics[width=0.45\linewidth]{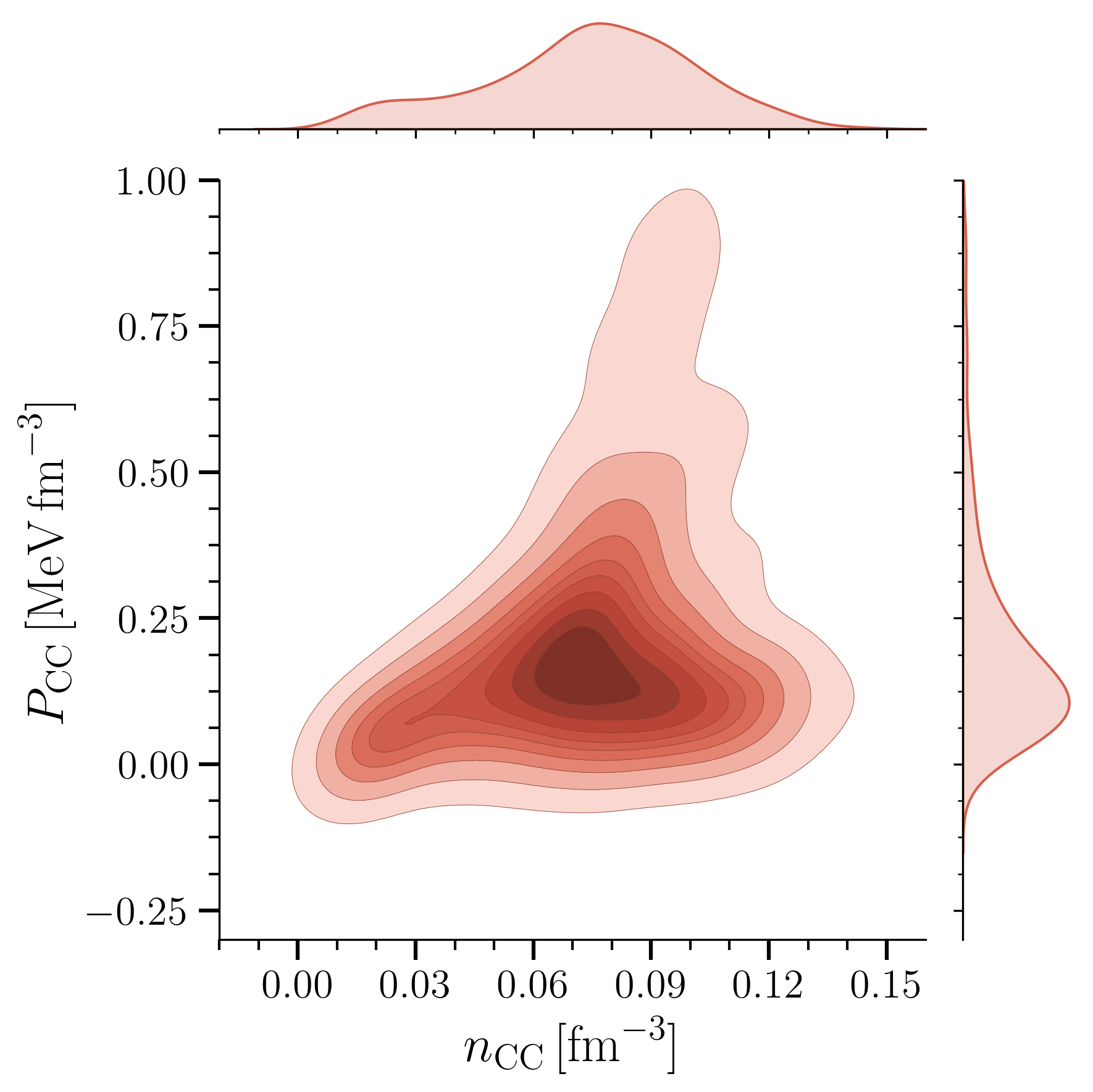}
		\label{fig:cc_MM_nochi}
	}
	\hfill
	\subfigure[MM (Post)]{
		\includegraphics[width=0.45\linewidth]{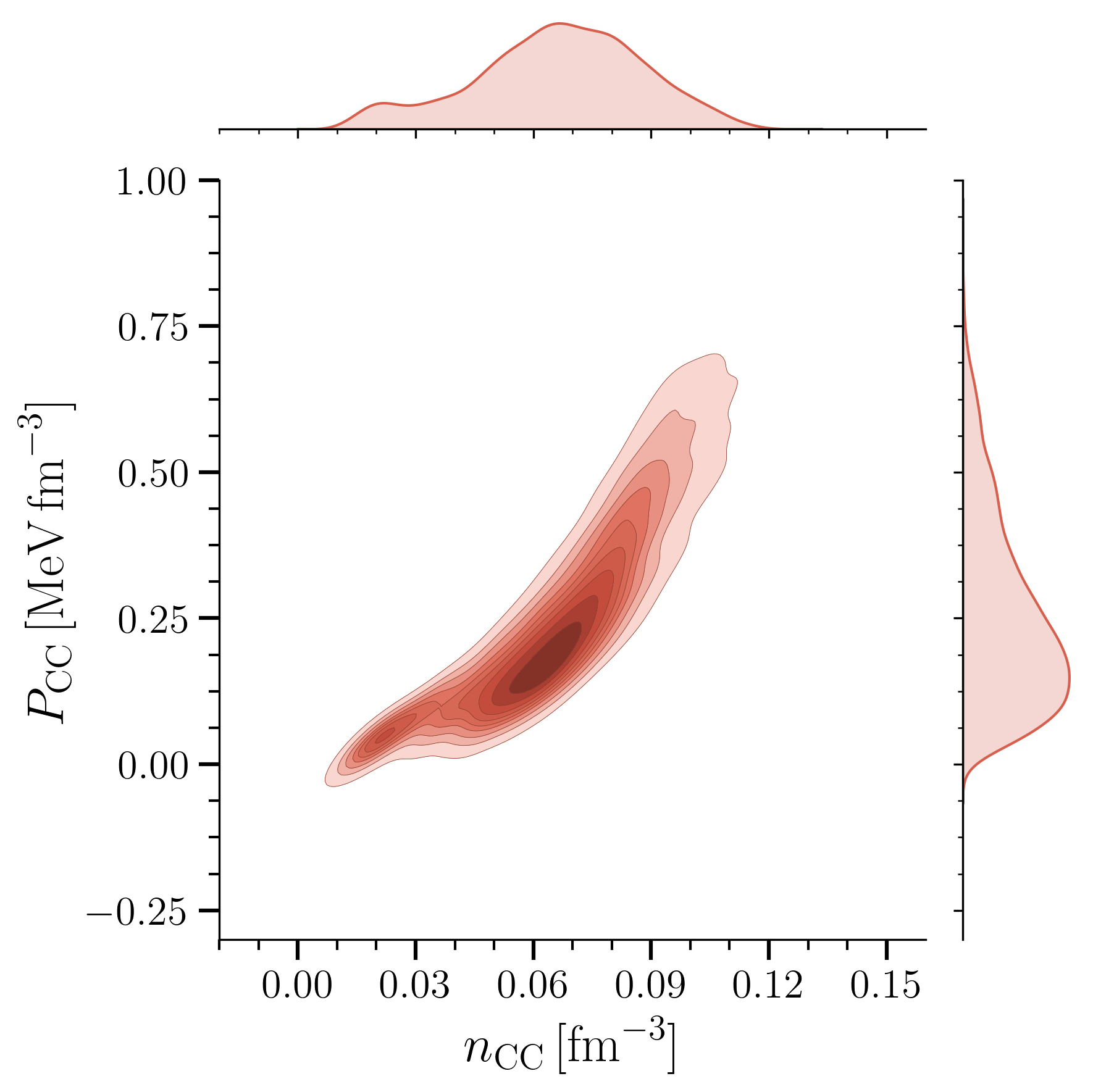}
		\label{fig:cc_MM_chi}
	}
	\hfill
	\subfigure[Y-MM (IP)]{
		\includegraphics[width=0.45\linewidth]{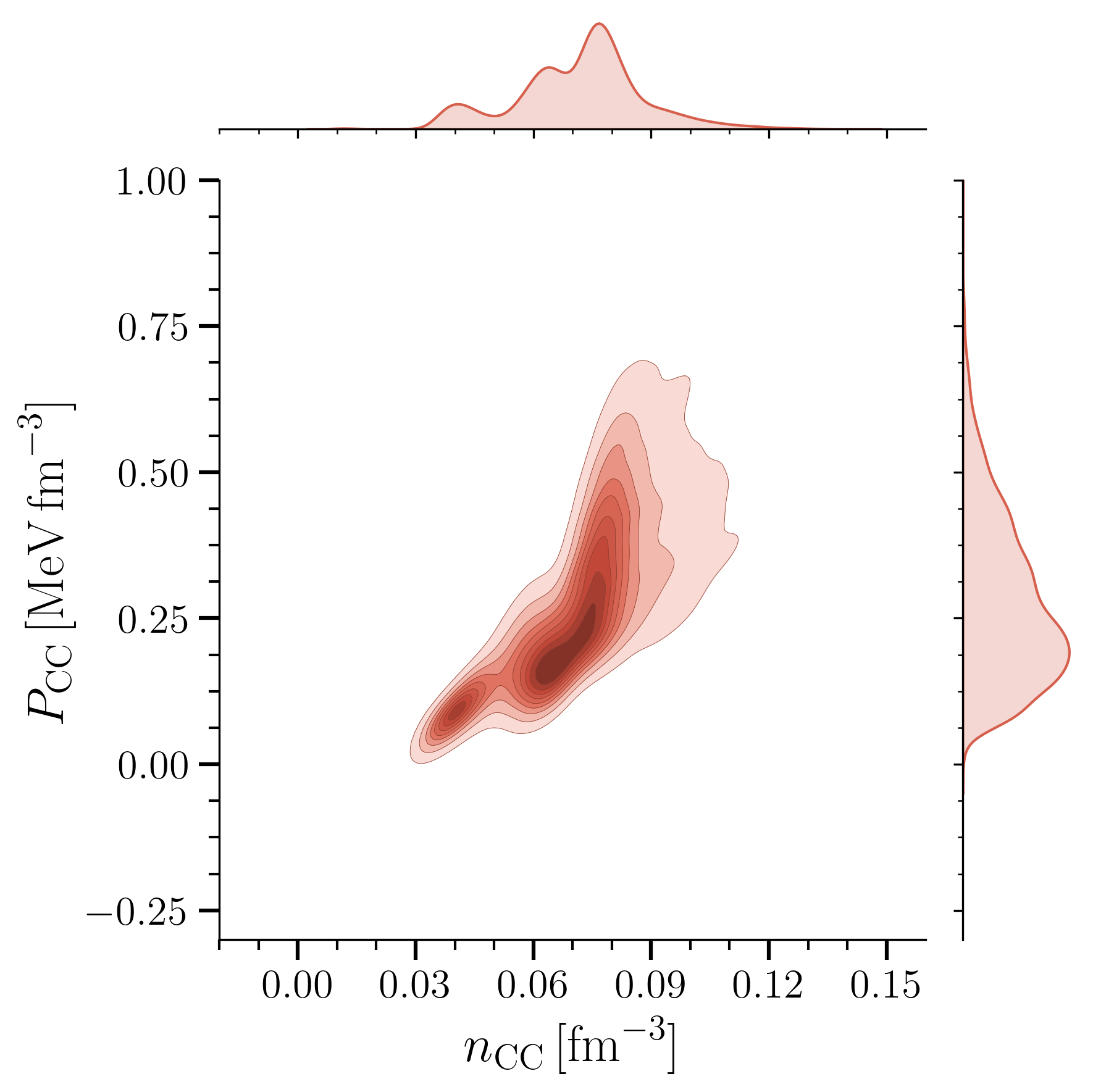}
		\label{fig:cc_MU_nochi}
	}
	\hfill
	\subfigure[Y-MM (Post)]{
		\includegraphics[width=0.45\linewidth]{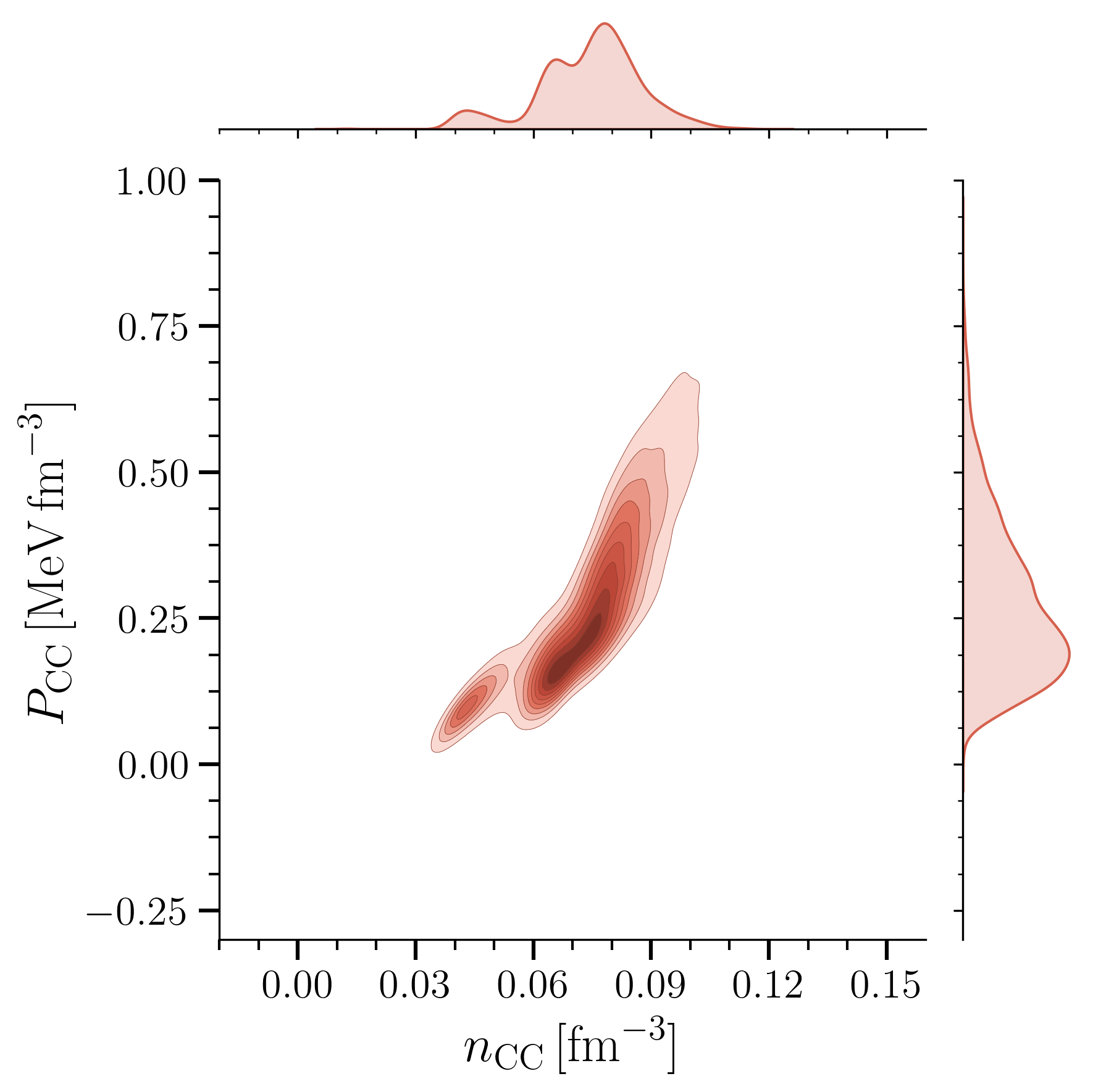}
		\label{fig:cc_MU_chi}
	}
	\caption{Joint probability density of the CC transition density $n_{\rm CC}$ and pressure $P_{\rm CC}$, obtained using the standard MM (top panels) and its \textit{ab-initio}-benchmarked version, Y-MM (bottom panels). The results from the IP distributions (left panels) are compared with those from the posterior distributions (right panels).}
	\label{fig:cc}
\end{figure}

Figure~\ref{fig:cc} shows the joint probability distributions of $n_{\rm CC}$ and $P_{\rm CC}$, along with the associated marginalized one-parameter distributions. Results are shown for the standard MM (top panels) and the \textit{ab-initio}-benchmarked Y-MM version (bottom panels). In each case, the IP distribution (left panels) is compared with the posterior distribution~(right panels).

From the top panels, it is evident that the IP filters alone do not strongly constrain the CC transition in the standard MM. In particular, the marginalized PDF for $P_{\rm CC}$ extends to very low values, and the distribution of $n_{\rm CC}$ remains broad (see panel (a) of Fig.~\ref{fig:cc}). As a result, the joint probability spans a wide region in the $(n_{\rm CC}, P_{\rm CC})$ plane. When the chiral filter is applied (panel (b)), the uncertainties are partially reduced, though the low-density tail in the $n_{\rm CC}$ distribution remains significantly populated.

By contrast, the bottom panels highlight the improvements introduced by the \textit{ab initio} benchmarking in the Y-MM approach. The joint distribution is notably more localized in the $(n_{\rm CC}, P_{\rm CC})$ diagram. Panels (c) and (d) are much more similar to each other than their MM counterparts, indicating that the chiral filter has limited impact in Y-MM: the nuclear-theory information is already embedded in the IP distribution. In both cases, the PDF for $P_{\rm CC}$ shifts toward higher values, while low transition densities in $n_{\rm CC}$ become strongly suppressed.

These findings are summarized in Table~\ref{tab:cc_density_pressure}. Beyond the overall reduction in the standard deviations of $n_{\rm CC}$ and $P_{\rm CC}$ predicted by Y-MM, the mean value of $n_{\rm CC}$ in the Y-MM posterior is larger than in the standard MM case, consistent with the increase in $L_{\rm sym}$ (see Table~\ref{tab:pdf_stats}), as previously noted in~\cite{TewsEPJA2019}. This also explains why the mean value of $n_{\rm CC}$ in MM (IP) nearly coincides with that of Y-MM (Post).
In contrast, the transition pressure $P_{\rm CC}$ shows a less regular dependence on $L_{\rm sym}$, being sensitive to correlations with higher order empirical parameters~\cite{CarreauEPJA2019}.

\begin{table}[tbp!]
	\centering
	\begin{tabular}{c|c|c}
		\toprule
		\bf{Model} & $n_{\rm{CC}}$ [fm$^{-3}$] & $P_{\rm{CC}}$ [MeV fm$^{-3}$]\\
		\midrule
		MM (IP) & 0.074 $\pm$ 0.027 & 0.201 $\pm$ 0.188 \\
		Y-MM (IP) & 0.072 $\pm$ 0.016 & 0.280 $\pm$ 0.143 \\
		MM (Post) & 0.066 $\pm$ 0.022 & 0.255 $\pm$ 0.161 \\  Y-MM (Post) & 0.074 $\pm$  0.014 & 0.277 $\pm$ 0.137 \\
		\bottomrule
	\end{tabular}
	\caption{Mean value and standard deviation of the crust-core transition density $n_{\rm CC}$ and pressure $P_{\rm CC}$, as obtained in the IP and posterior distributions of the standard MM and its \textit{ab-initio}-benchmarked version Y-MM.}
	\label{tab:cc_density_pressure}
\end{table}

\subsection{Inference of crustal properties: moment of inertia}

As a final application, we compute the fractional moment of inertia of the crust, $\mathcal{I}_{\rm crust}/\mathcal{I}$, within the slow-rotation approximation~\cite{DelsatePRD2016,ant2018MNRAS}. This quantity is key to interpreting the average glitch activity of the Vela pulsar (PSR J0835–4510), which requires a substantial angular momentum reservoir in the crustal superfluid~\cite{LinkPRL1999,Chamel2012,AnderssonPRL2012}; see~\cite{amp_review_2023} for a review.
The connection between inner-crust structure and glitch activity is expressed by the inequality~\cite{amp_review_2023}
\begin{equation}
	\frac{\mathcal{I}_n}{\mathcal{I} - \mathcal{I}_n} > \mathcal{G} \,,
	\label{eq:glitch_activity_crust}
\end{equation}
where $\mathcal{I}$ is the total moment of inertia, $\mathcal{I}_n$ is the moment of inertia of the superfluid neutrons in the inner crust, and $\mathcal{G}$ is the dimensionless glitch activity inferred from long-term pulsar timing. To date, the most active glitching pulsar is Vela, for which a heteroscedastic fit yields~\cite{Montoli2021}
\begin{equation}
    \mathcal{G}_{\text{Vela}} \approx 0.016 \pm 0.002 \,.
\end{equation}
The calculation of $\mathcal{I}_n$ depends on microscopic estimates of the superfluid density, which can differ from the total neutron density due to entrainment: a non-dissipative coupling to the crustal lattice that reduces the mobility of neutrons~\cite{Chamel2008,amp_review_2023}. However, a recent revision of entrainment calculations by~\citet{Almirante2025arXiv} suggests that the effect may be significantly weaker than previously estimated. This justifies the approximation $\mathcal{I}_n \approx \mathcal{I}_{\rm crust}$. Since $\mathcal{I}_{\rm crust} \ll \mathcal{I}$, Eq.~\eqref{eq:glitch_activity_crust} reduces to the original zero-entrainment condition~\cite{LinkPRL1999}:
\begin{equation}
	\frac{\mathcal{I}_{\rm crust}}{\mathcal{I}} > \mathcal{G}_{\text{Vela}} \,.
\end{equation}
Since Vela is the pulsar with the largest well-determined value of $\mathcal{G}$~\cite{Montoli2021}, this inequality sets an observational lower bound on the fractional moment of inertia of the crust.

\begin{figure}[tbp!]
	\centering
	\subfigure[MM (IP)]{
		\includegraphics[width=0.45\linewidth]{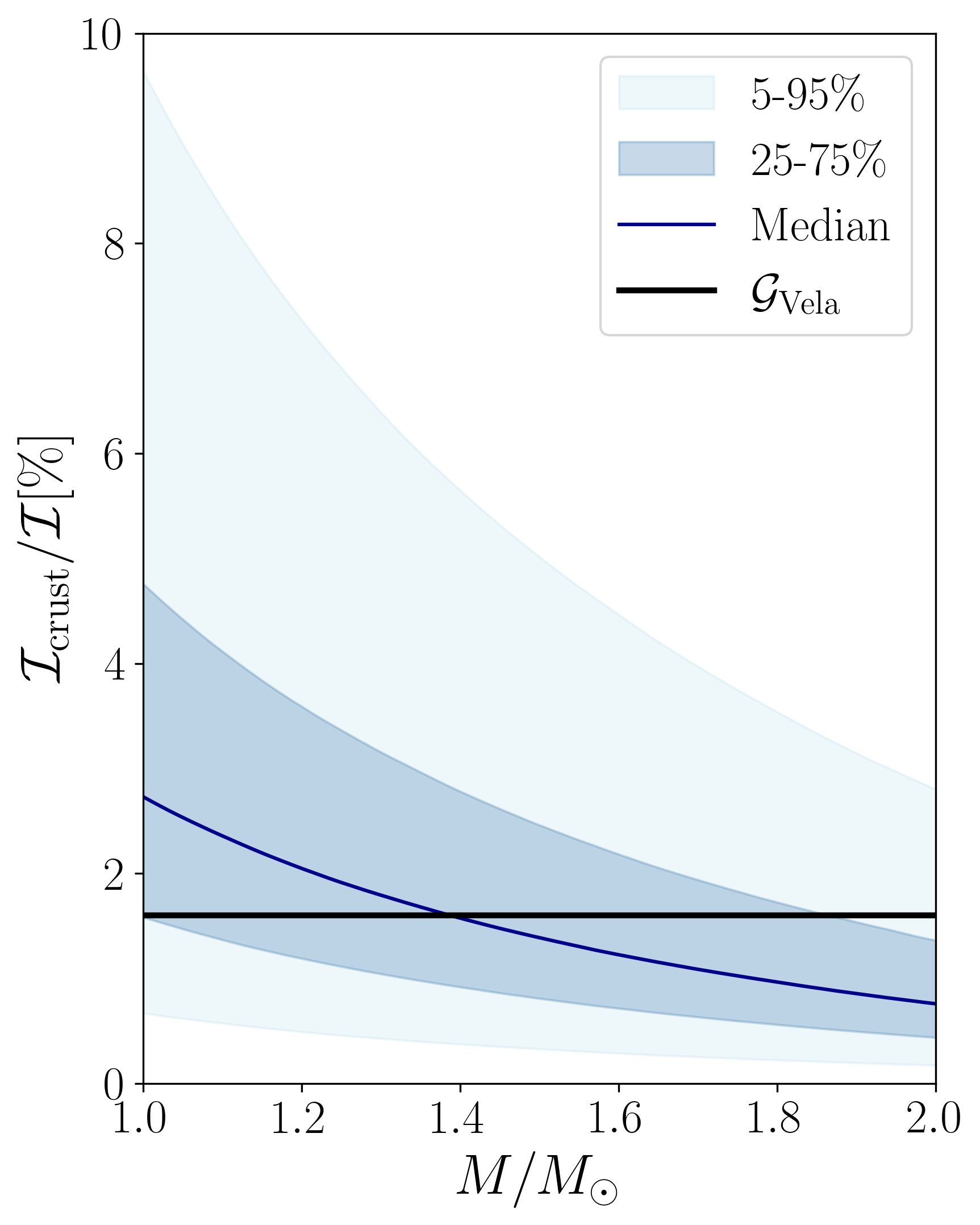}
		\label{fig:inertia_MM_nochi}
	}
	\hfill
	\subfigure[MM (Post)]{
		\includegraphics[width=0.45\linewidth]{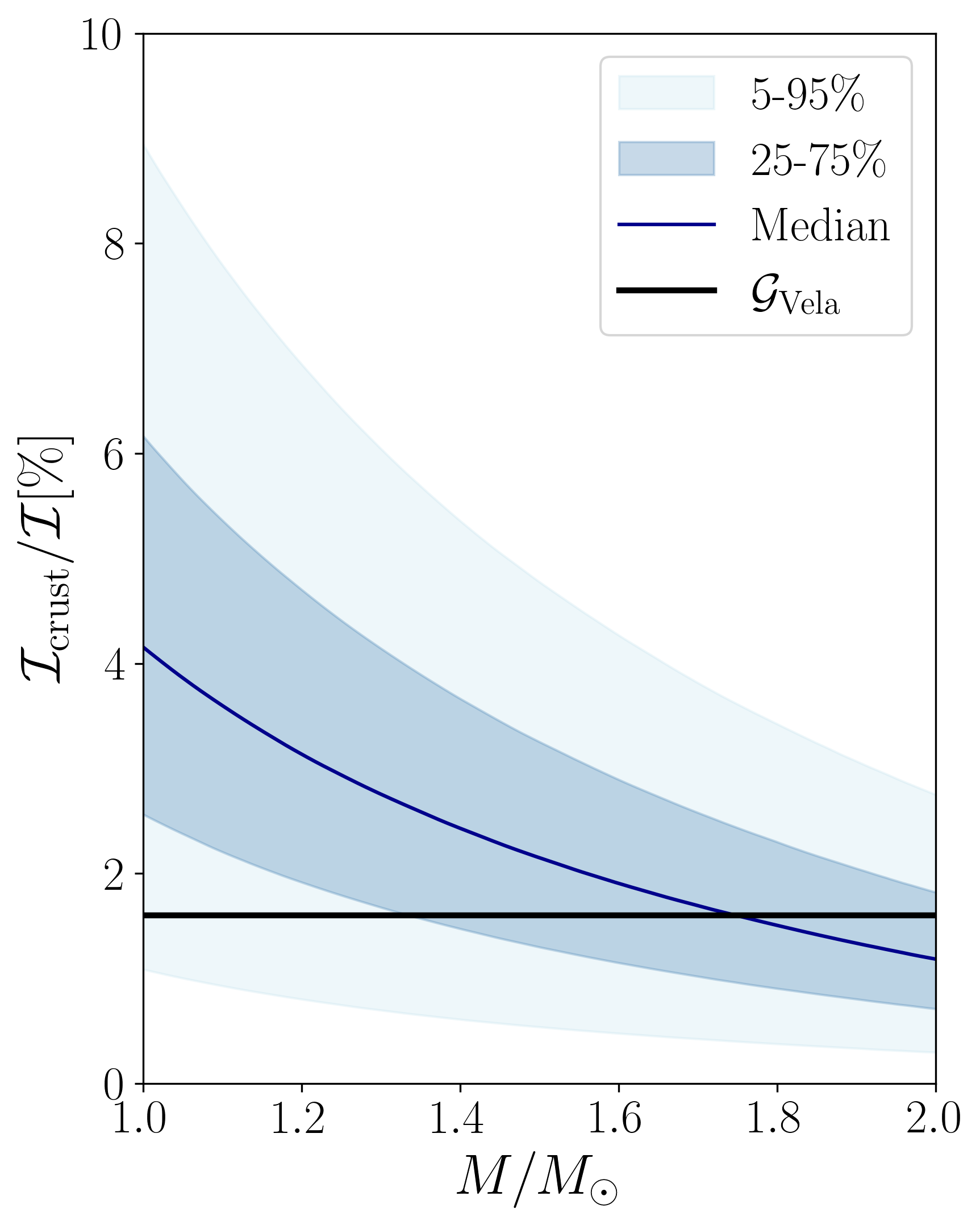}
		\label{fig:inertia_MM_chi}
	}
	\hfill
	\subfigure[Y-MM (IP)]{
		\includegraphics[width=0.45\linewidth]{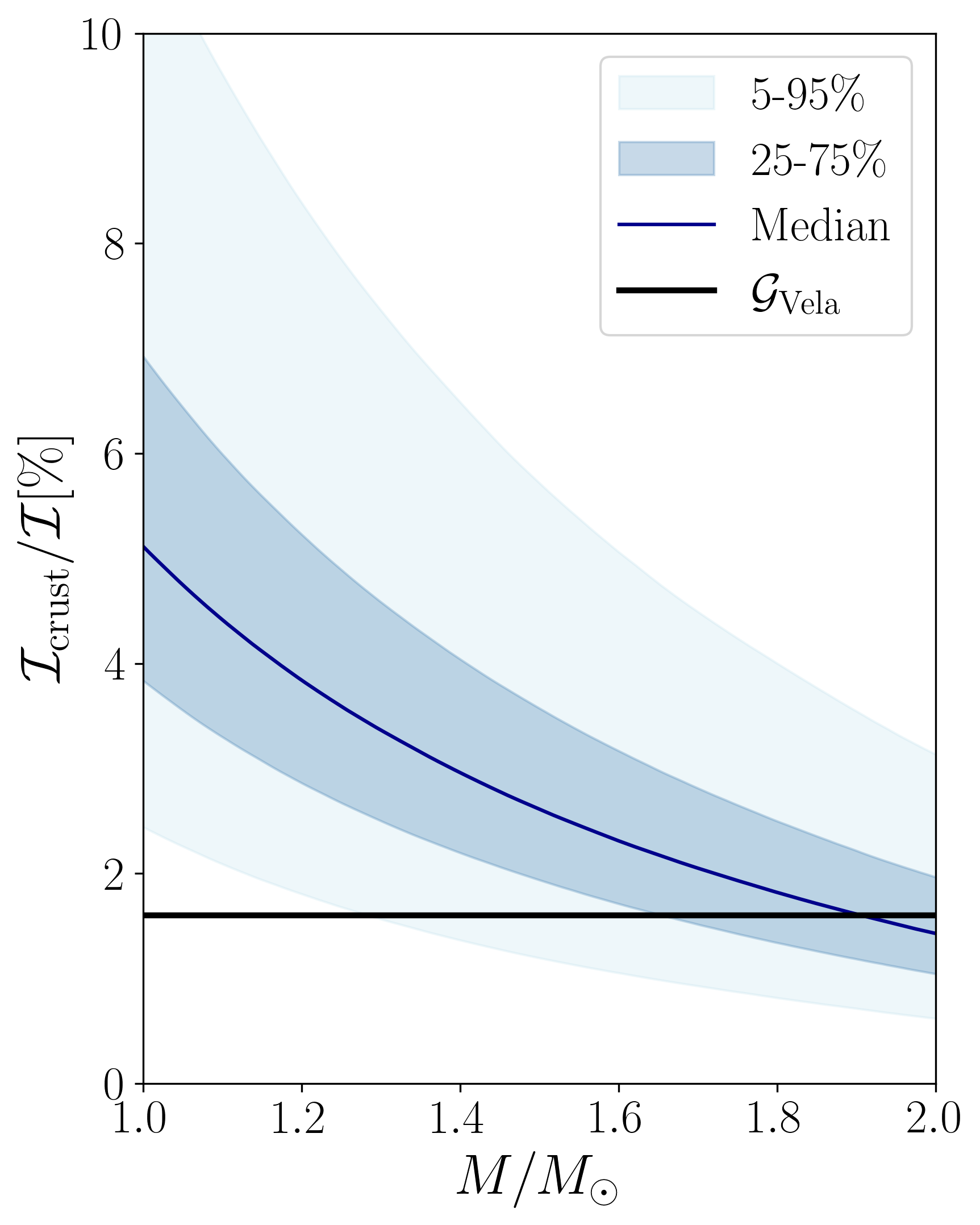}
		\label{fig:inertia_MU_nochi}
	}
	\hfill
	\subfigure[Y-MM (Post)]{
		\includegraphics[width=0.45\linewidth]{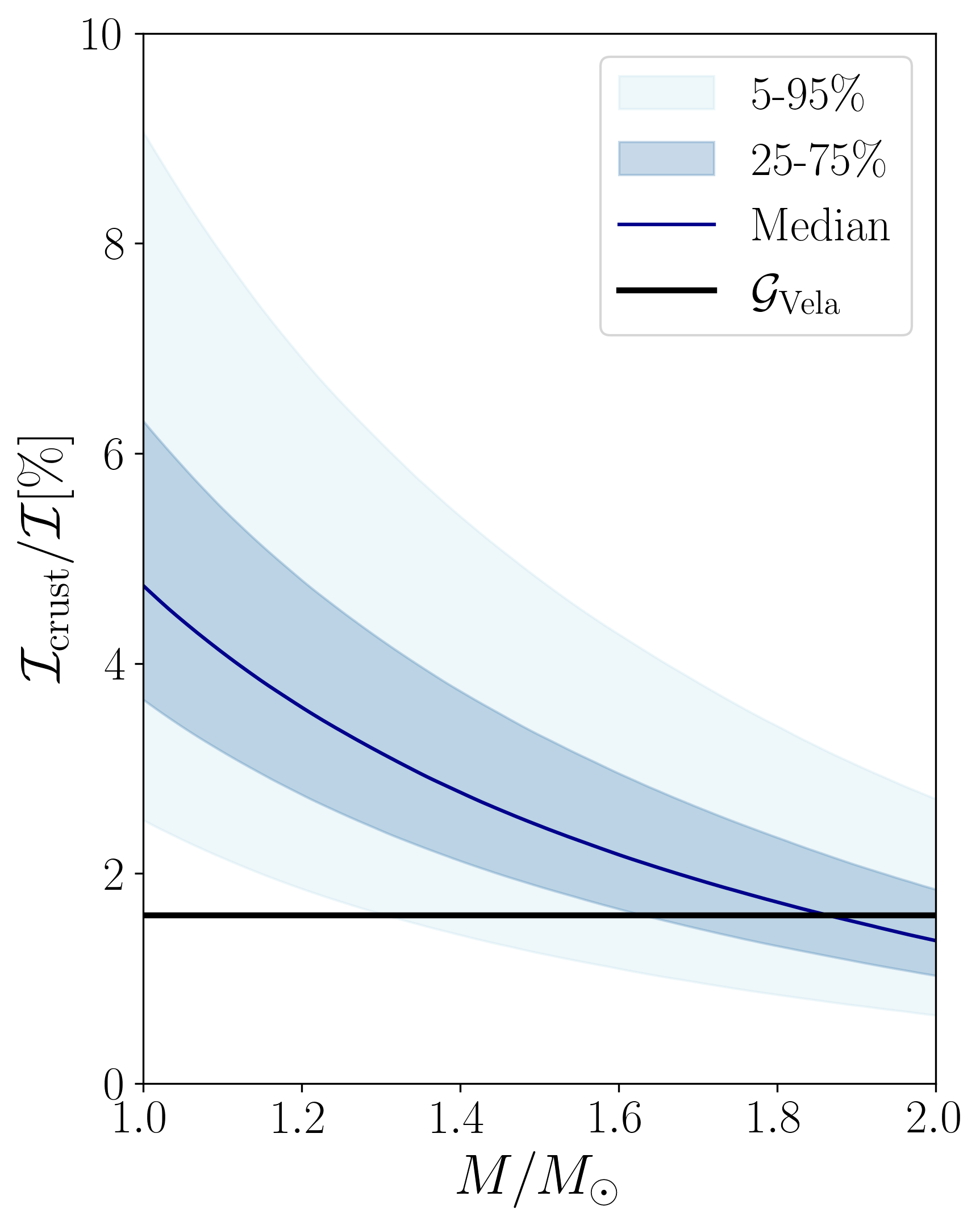}
		\label{fig:inertia_MU_chi}
	}
	\caption{Percentage of the crustal fraction of the moment of inertia $\mathcal{I}_{\rm crust}/\mathcal{I}$ as a function of the NS mass $M/M_{\odot}$, as obtained within the standard MM (top panels) or its \textit{ab-initio}-benchmarked Y-MM version (bottom panels). The predictions from the IP distributions (left panels) are compared with the posterior ones (right panels). The dark and light blue shaded regions represent the 50\% and 90\% confidence intervals, respectively. The dark blue line indicates the median value and the horizontal black line refers to the value of $\mathcal{G}_{\text{Vela}} \approx 1.6\%$ as inferred with the heteroscedastic fit proposed in~\cite{Montoli2021}. }
	\label{fig:inertia}
\end{figure}

In Fig.~\ref{fig:inertia}, we plot $\mathcal{I}_{\rm crust}/\mathcal{I}$ as a function of stellar mass $M$, normalized to the solar mass $M_{\odot}$. As in Fig.~\ref{fig:cc}, we compare the standard MM approach (top panels) with the \textit{ab-initio}-benchmarked Y-MM method (bottom panels). The effect of the chiral filter is shown by comparing IP distributions (left panels) with the full posteriors (right panels).

In the standard MM case, the IP filters weakly constrain the observable, especially for lower-mass stars (panel (a)) and the chiral filter is needed not only to reduce the upper tail, but also to shift upward the $\mathcal{I}_{\rm crust}/\mathcal{I}$ posterior distributions. 
By contrast, the Y-MM predictions are already well-constrained at the IP level (panel (c)) and systematically shifted to higher values. Notably, Y-MM predictions satisfy the inequality $\mathcal{I}_{\rm crust}/\mathcal{I} > \mathcal{G}_{\text{Vela}}$ for essentially all plausible Vela masses, with the exception of stars heavier than $\sim 1.8\,M_\odot$.
This supports the interpretation that Vela glitches can be attributed to vortex pinning in the crust, assuming entrainment is negligible~\cite{amp_review_2023}. However, if entrainment were significantly stronger, the required moment of inertia would increase, potentially ruling out a purely crustal origin of glitches~\cite{Chamel2012,AnderssonPRL2012,DelsatePRD2016,Montoli2021}.

Given the limited constraining power of $\mathcal{G}_{\text{Vela}}$ under the conservative assumption of negligible entrainment, our results show that crustal moment of inertia predictions remain robust across models and filters, and are compatible with the observational constraint imposed by Vela, in agreement with the recent Bayesian analysis of~\cite{klausner2025arXiv}.
 
\section{Conclusions}
\label{sec:conclusions}

We have extended the unified MM framework for the nuclear EoS by introducing low-density corrections based on EDFs benchmarked against \textit{ab initio} PNM calculations. This upgrade addresses a long-standing limitation of phenomenological models in the dilute regime, enhancing consistency with microscopic PNM physics near the unitary limit.
Although the influence of low-density corrections remains secondary to core modeling uncertainties, our results emphasize the importance of incorporating realistic low-density physics into unified EoS models.  The refined treatment in the dilute regime causes the Y-MM predictions to diverge from those of the standard MM, as the latter fails to reproduce nuclear theory constraints in the zero-density limit (below 0.02 fm$^{-3}$).

Implementing these corrections within the Y-MM approach, we assessed their impact on crustal properties, within a Bayesian inference framework. The introduced correction reduces the model dependence that affects estimates of crustal composition and CC transition.
Beyond the overall reduction in standard deviations predicted by Y-MM, our analysis reveals sizable effects on several key observables. In particular, the charge of nuclear clusters in the deep inner crust decreases, potentially impacting transport properties.

%\ora{FG comment: the EoS is not always softened...the reason is not fully understood}
%\cya{SB: To a large extent, the value of $e_{\rm B}$ for PNM with Y-MM lies close to the lower edge of the chiral band, slightly reducing the stiffness of the EoS at sub-saturation densities and thereby inducing a mild stiffening at higher densities (around saturation and beyond).}
Moreover, the Y-MM informed prior corresponds to a systematic shift in the distribution of the symmetry-energy slope parameter $L_{\text{sym}}$ toward stiffer behavior. 
This can be understood from the fact that the controlled low-density limit can be matched to an arbitrary EoS keeping stability and thermodynamic consistency, only if this latter is sufficiently stiff around saturation. Even if the effect is mitigated by the application of the chiral constraint, this, in turn, leads to an increase in the CC transition density and pressure, as well as a higher fraction of the crustal moment of inertia. The latter supports interpreting Vela glitches as the result of vortex pinning in the crust—assuming negligible entrainment—across a somewhat broader range of stellar masses when Y-MM is used instead of~MM. While the present work focuses on crustal properties, we note that models fulfilling the maximum mass constraint predict, for a \(1.4\,M_{\odot}\) NS, radii in the range of 11–14 km, which is broadly consistent with current NICER measurements~\cite{Miller2021, Riley2021}; a more detailed analysis of radius predictions will be presented in future studies.

A practical strength of the blending procedure, Eq.~\eqref{eq:interpolation}, is that it can be applied to any EoS framework at negligible computational cost, including relativistic mean-field models, which typically describe high-density matter well but often diverge from \textit{ab initio} predictions in the dilute regime. The Y-MM framework thus offers a coherent modeling of NS matter across all densities, providing an avenue to assess EoS-related uncertainties in crust modeling, particularly when incorporated into a Bayesian setup.

\begin{acknowledgements}
	Stimulating discussions with Marcella Grasso and Isaac Vida\~na are gratefully acknowledged. The authors also thank Michael Urban and Viswanathan Palaniappan for providing the BMBPT3 benchmark data, and Gabriele Montefusco for helpful coding advice.
	F.G., A.F.F., and M.A. acknowledge the support by the IN2P3 Master Project NewMAC and MAC, the ANR project `Gravitational waves from hot neutron stars and properties of ultra-dense matter' (GW-HNS, ANR-22-CE31-0001-01), and the CNRS International Research Project (IRP) `Origine des \'el\'ements lourds dans l'univers: Astres Compacts et Nucl\'eosynth\`ese (ACNu)'.
\end{acknowledgements}

\newpage
\appendix
\begin{widetext}
	
\section{Density and asymmetry derivatives of the Fermi gas energy}
\label{app:derivatives_fg}

The non-relativistic FG energy $t_{\rm FG}^{\ast}$, may be written as
\begin{equation}
	t_{\rm FG}^{\ast} (n_{\rm B}, \delta) = \dfrac{3}{5 n_{\rm B}} \left( \varepsilon_{\rm F,n}^{\ast}n_{\rm n} + \varepsilon_{\rm F,p}^{\ast}n_{\rm p}\right) \ ,
	\label{eq:tfg2}
\end{equation}
where 
\begin{equation}
	\varepsilon_{\rm F,q}^{\ast} =\dfrac{\hbar^{2}}{2 m_{\rm q}^{\ast}} \left(3\pi^{2}n_{\rm q}\right)^{2/3}, \qquad {\rm q = n, p}
\end{equation}
is the Fermi energy of the generic species ${\rm q}$ considering the momentum dependence of the nuclear interaction. 
Using the definition of effective mass, as given in Eq.~(10) of~\cite{MargueronPRC2018}:
\begin{equation}
	m_{\rm q}^{\ast} = \dfrac{m \, n_{\rm sat}}{n_{\rm sat}  + \left( \kappa_{\rm sat} + \tau_{3} \kappa_{\rm sym} \delta \right) n_{\rm B}  }\ ,
\end{equation}
with $\tau_{3} = \pm 1$, for ${\rm n}$ and ${\rm p}$, respectively, and the definition of neutron and proton densities
\begin{equation}
	n_{\rm n} = \dfrac{n_{\rm B}}{2} \left(1 + \delta \right) \ , \qquad  n_{\rm p} = \dfrac{n_{\rm B}}{2} \left(1 - \delta \right) \ ,    
\end{equation}
one gets
\begin{eqnarray}
	t_{\rm FG}^{\ast} (n_{\rm B}, \delta) 
	& = & \dfrac{t_{\rm FG, sat}}{2} \left( \dfrac{n_{\rm B}}{n_{\rm sat}} \right)^{2/3} \left[ \left( 1 + \kappa_{\rm sat} \dfrac{n_{\rm B}}{n_{\rm sat}} \right) f_{1} (\delta) + \kappa_{\rm sym} \dfrac{n_{\rm B}}{n_{\rm sat}} f_{2} (\delta) \right] \ , \nonumber
\end{eqnarray}
which is exactly Eq.~\eqref{eq:tfg}.

The density and isospin asymmetry of the FG energy can be then easily derived from Eq.~\eqref{eq:tfg2}. For that purpose, let us firstly derive the derivatives of the nucleon effective mass with respect to the density 
\begin{eqnarray}
	\dfrac{\partial m_{\rm q}^{\ast}}{\partial n_{\rm B}} & = &  \dfrac{\partial}{\partial n_{\rm B}} \left [\dfrac{m}{1 + \left( \kappa_{\rm sat} + \tau_{3} \kappa_{\rm sym} \delta \right) \dfrac{n_{\rm B}}{n_{\rm sat}}} \right]  
	= - \dfrac{\left(m_{\rm q}^{\ast}\right)^{2}}{m}\left( \dfrac{\kappa_{\rm sat} + \tau_{3} \kappa_{\rm sym} \delta }{n_{\rm sat}} \right)  
\end{eqnarray}
and isospin asymmetry
\begin{eqnarray}
	\dfrac{\partial m_{\rm q}^{\ast} }{\partial \delta} 
	& = &  \dfrac{\partial}{\partial \delta} \left [\dfrac{m}{1 + \left( \kappa_{\rm sat} + \tau_{3} \kappa_{\rm sym} \delta \right) \dfrac{n_{\rm B}}{n_{\rm sat}}} \right] 
	= - \dfrac{\left(m_{\rm q}^{\ast}\right)^{2}}{m}\left( \tau_{3} \kappa_{\rm sym} \dfrac{ n_{\rm B} }{n_{\rm sat}} \right),
\end{eqnarray}
respectively.

The quantities derived above enter into the derivatives of the Fermi energy, with respect to density
\begin{eqnarray}
	\dfrac{\partial \varepsilon_{\rm F,q}^{\ast}}{\partial n_{\rm B}} & = & 
	\dfrac{\varepsilon_{\rm F,q}^{\ast}}{n_{\rm B}} \left[ \dfrac{m_{\rm q}^{\ast}}{m} \left( \kappa_{\rm sat} + \tau_{3} \kappa_{\rm sym} \delta \right) \left( \dfrac{n_{\rm B}}{n_{\rm sat}} \right)
	+ \dfrac{2}{3} \right]
\end{eqnarray}
and isospin asymmetry
\begin{eqnarray}    
	\dfrac{\partial \varepsilon_{\rm F,q}^{\ast}}{\partial \delta} & = & 
	\tau_{3}\varepsilon_{\rm F,q}^{\ast}  \left( \dfrac{m_{\rm q}^{\ast}}{m} \kappa_{\rm sym} \dfrac{ n_{\rm B} }{n_{\rm sat}} +\dfrac{1}{1 + \tau_{3}\delta} \dfrac{2}{3}  \right) \ ,  \nonumber \\
\end{eqnarray}
respectively.
In case of a momentum-independent interaction, as for the YGLO EDF, one has
\begin{equation}
	\dfrac{\partial \varepsilon_{\rm F,q}}{\partial n_{\rm B}} = \dfrac{2}{3} \dfrac{\varepsilon_{\rm F,q}}{n_{\rm B}} 
\end{equation}
and
\begin{equation}    
	\dfrac{\partial \varepsilon_{\rm F,q}}{\partial \delta} = \dfrac{\tau_{3}}{1 + \tau_{3}\delta} \dfrac{2}{3} \varepsilon_{\rm F,q} \ .
\end{equation}

\subsubsection{Density and asymmetry derivatives of potential energy}

Let us derive the derivatives of the potential energy per nucleon, with respect to the baryon number density and isospin asymmetry. For the standard MM, the potential energy per nucleon writes
\begin{equation}
	v_{\rm MM} = \sum_{\alpha = 0}^{4} \dfrac{1}{\alpha!} \left(v_{\alpha}^{\rm is} + v_{\alpha}^{\rm iv} \delta^{2}  \right) x^{\alpha} \ ,
\end{equation}
where, with respect to Eq.~\eqref{eq:vn}, we have taken $\mathcal{N} = 4$. Then
\begin{equation}
	\dfrac{\partial v_{\rm MM} (n_{\rm B}, \delta)}{\partial n_{\rm B}}  = \sum_{\alpha = 1}^{4} \dfrac{1}{\left(\alpha-1\right)!} \left(v_{\alpha}^{\rm is} + v_{\alpha}^{\rm iv} \delta^{2}  \right) x^{\alpha-1}   
\end{equation}
and
\begin{equation}
	\dfrac{\partial v_{\rm MM} (n_{\rm B}, \delta)}{\partial \delta} = 2 \delta  \sum_{\alpha = 0}^{4} \dfrac{1}{\alpha!}  v_{\alpha}^{\rm iv} x^{\alpha}   \ .
\end{equation}

On the other hand, for the YGLO functional the derivatives write as
\begin{eqnarray}
	\dfrac{\partial v_{\rm Y} (n_{\rm B}, \delta)}{\partial n_{\rm B}} & = & \dfrac{\partial}{\partial n_{\rm B}} \left\{ \dfrac{1}{n_{\rm B}} \left[ \mathcal{V}_{\rm s}^{\rm Y} + \left(\mathcal{V}_{\rm n}^{\rm Y} -\mathcal{V}_{\rm s}^{\rm Y} \right) \delta^{2} \right] \right\} \nonumber \\
	& = & \dfrac{\partial}{\partial n_{\rm B}} \left[ Y_{\rm s}  n_{\rm B} + D_{\rm s} n_{\rm B}^{5/3} + F_{\rm s} n_{\rm B}^{(\alpha + 1)} + \left( Y_{\rm n}  n_{\rm B} + D_{\rm n} n_{\rm B}^{5/3} + F_{\rm n} n_{\rm B}^{(\alpha + 1)} - Y_{s}  n_{\rm B} - D_{\rm s} n_{\rm B}^{5/3} - F_{\rm s} n_{\rm B}^{(\alpha + 1)}\right) \delta^{2} \right ] \nonumber \\ 
	&& + \left[ \dfrac{\partial Y_{\rm n}}{\partial n_{\rm B}}n_{\rm B} + Y_{\rm n} + \dfrac{5}{3}D_{\rm n} n_{\rm B}^{2/3} + (\alpha + 1) F_{\rm n} n_{\rm B}^{\alpha} - \dfrac{\partial Y_{\rm s}}{\partial n_{\rm B}}n_{\rm B} - Y_{\rm s} - \dfrac{5}{3}D_{\rm s} n_{\rm B}^{2/3} - (\alpha + 1) F_{\rm s} n_{\rm B}^{\alpha} \right] \delta^{2} \nonumber \\ 
\end{eqnarray}
and
\begin{eqnarray}
	\dfrac{\partial v_{\rm Y} (n_{\rm B}, \delta)}{\partial \delta} & = & \dfrac{\partial}{\partial \delta} \left\{ \dfrac{1}{n_{\rm B}} \left[ \mathcal{V}_{\rm s}^{\rm Y} + \left(\mathcal{V}_{\rm n}^{\rm Y} -\mathcal{V}_{\rm s}^{\rm Y} \right) \delta^{2} \right] \right\} \nonumber \\
	& = & 2 \delta \left( Y_{\rm n}  n_{\rm B} + D_{\rm n} n_{\rm B}^{5/3} + F_{\rm n} n_{\rm B}^{(\alpha + 1)} - Y_{s}  n_{\rm B} - D_{\rm s} n_{\rm B}^{5/3} - F_{\rm s} n_{\rm B}^{(\alpha + 1)}\right) \ ,
\end{eqnarray}
where
\begin{eqnarray}
	\dfrac{\partial Y_{\rm i}}{\partial n_{\rm B}} & = & \dfrac{\partial} {\partial n_{\rm B}} \left ( \frac{B_{\rm i}}{1 - R_{\rm i} n_{\rm B}^{1/3} + C_{\rm i} n_{\rm B}^{2/3}} \right) 
	= \dfrac{Y_{\rm i}^{2}}{3 B_{\rm i} n_{\rm B}} \left( R_{\rm i} n_{\rm B}^{1/3} - 2C_{\rm i} n_{\rm B}^{2/3} \right), \qquad {\rm i = s, n} \ .
\end{eqnarray}

\section{Constraining the transition function to avoid spurious PNM instabilities}
\label{app:inequality}

To avoid the emergence of (spurious) PNM spinodal instabilities in the chemical potential derivatives, we must impose that, in case of PNM, the follow inequality, 
\begin{equation}
	\dfrac{\partial \mu_{\rm n}}{\partial n_{\rm B}} \ge 0     \ ,
\end{equation}
applies  everywhere, or, equivalently, as in Eq.~\eqref{eq:inequality} that
\begin{eqnarray}
	\dfrac{\partial^{2} e_{\rm B}}{\partial n_{\rm B}^{2}} + \dfrac{2}{n_{\rm B}} \dfrac{\partial e_{\rm B}}{\partial n_{\rm B}} \ge 0 \ .
\end{eqnarray}
Let us explicit the density derivatives which appear in Eq.~\eqref{eq:inequality} that, according to  Eq.~\eqref{eq:interpolation},
are:
\begin{eqnarray}
	\dfrac{\partial e_{\rm B}}{\partial n_{\rm B}} 
	& = &  \left(e_{\rm MM} - e_{\rm Y} \right) \dfrac{\partial \eta_{\chi}^{\rm MM}}{\partial n_{\rm B}} + \left[ \dfrac{\partial e_{\rm Y}}{\partial n_{\rm B}} \left( 1 - \eta_{\chi}^{\rm MM} \right) + \dfrac{\partial e_{\rm MM}}{\partial n_{\rm B}}\eta_{\chi}^{\rm MM} \right] 
\end{eqnarray}
and
\begin{eqnarray}
	\dfrac{\partial^{2} e_{\rm B}}{\partial n_{\rm B}^{2}} 
	& = & \left(e_{\rm MM} - e_{\rm Y} \right) \dfrac{\partial^{2} \eta_{\chi}^{\rm MM}}{\partial n_{\rm B}^{2}} + 2 \left( \dfrac{\partial e_{\rm MM}}{\partial n_{\rm B}} - \dfrac{\partial e_{\rm Y}}{\partial n_{\rm B}} \right)  \dfrac{\partial \eta_{\chi}^{\rm MM}}{\partial n_{\rm B}}  + \left[ \dfrac{\partial^{2} e_{\rm Y}}{\partial n_{\rm B}^{2}} \left( 1 - \eta_{\chi}^{\rm MM} \right) + \dfrac{\partial^{2} e_{\rm MM}}{\partial n_{\rm B}^{2}}\eta_{\chi}^{\rm MM} \right] \ .
\end{eqnarray}

Finally, the inequality expressed by Eq.~\eqref{eq:inequality} can be written in the following compact form:
\begin{eqnarray}
	\left(e_{\rm MM} - e_{\rm Y} \right) \dfrac{\partial^{2} \eta_{\chi}^{\rm MM}}{\partial n_{\rm B}^{2}} + \dfrac{2}{n_{\rm B}} \left( \mu_{\rm n, MM} - \mu_{\rm n, Y}\right) \dfrac{\partial \eta_{\chi}^{\rm MM}}{\partial n_{\rm B}} + \dfrac{1}{n_{\rm B}} \left[ \dfrac{\partial \mu_{\rm n, MM}}{\partial n_{\rm B}} \eta_{\chi}^{\rm MM}  +  \dfrac{\partial \mu_{\rm n, Y}}{\partial n_{\rm B}} \left( 1 - \eta_{\chi}^{\rm MM} \right) \right] \ge 0  \ .
\end{eqnarray}

To check the inequality in Eq.~\eqref{eq:inequality}, let us first write explicitly the quantities there involved, that are the first-order density derivative,
\begin{equation}
	\dfrac{\partial \eta_{\chi}^{\rm MM}}{\partial n_{\rm B}} =
	\dfrac{f^{\prime} (x_{\chi}^{\rm MM}) f (1 - x_{\chi}^{\rm MM}) + f (x_{\chi}^{\rm MM}) f^{\prime}  (1 - x_{\chi}^{\rm MM})}{\left[ f(x_{\chi}^{\rm MM}) + f (1 - x_{\chi}^{\rm MM}) \right]^{2}} \dfrac{\partial x_{\chi}^{\rm MM}}{\partial n_{\rm B}} \ , \label{eq:deta}
\end{equation} 
and the second-order density derivative of the smoothing function introduced by Eq.~\eqref{eq:mollifier},
\begin{eqnarray}
	\dfrac{\partial^{2} \eta_{\chi}^{\rm MM}}{\partial n_{\rm B}^{2}} & = & \dfrac{f^{\prime\prime} (x_{\chi}^{\rm MM}) f (1 - x_{\chi}^{\rm MM}) - f (x_{\chi}^{\rm MM}) f (1 - x_{\chi}^{\rm MM}) }{\left[f (x_{\chi}^{\rm MM}) + f (1 - x_{\chi}^{\rm MM}) \right]^{2}} \left(\dfrac{\partial x_{\chi}^{\rm MM}}{\partial n_{\rm B}} \right)^{2} \nonumber \\
	&& - 2 \dfrac{\left[ f^{\prime} (x_{\chi}^{\rm MM}) - f^{\prime} (1 - x_{\chi}^{\rm MM})  \right] \left[ f^{\prime} (x_{\chi}^{\rm MM}) f (1 - x_{\chi}^{\rm MM}) + f (x_{\chi}^{\rm MM}) f^{\prime} (1 - x_{\chi}^{\rm MM}) \right]}{\left[f (x_{\chi}^{\rm MM}) + f (1 - x_{\chi}^{\rm MM}) \right]^{3}} \left(\dfrac{\partial x_{\chi}^{\rm MM}}{\partial n_{\rm B}} \right)^{2} \label{eq:d2eta},
\end{eqnarray}
the neutron chemical potentials
\begin{eqnarray}
	\mu_{\rm n, MM} & = & \dfrac{3}{5} \varepsilon_{\rm F, n}^{\ast} + \dfrac{3}{5} n_{\rm B} \dfrac{\partial \varepsilon_{\rm F, n}^{\ast}}{\partial n_{\rm B}} + v_{\rm MM} +  n_{\rm B} \dfrac{\partial v_{\rm MM}}{\partial n_{\rm B}} \ , \nonumber \\
	\mu_{\rm n, Y} & = &\dfrac{3}{5} \varepsilon_{\rm F, n}+ \dfrac{3}{5} n_{\rm B} \dfrac{\partial \varepsilon_{\rm F, n}}{\partial n_{\rm B}} + v_{\rm Y} +  n_{\rm B} \dfrac{\partial v_{\rm Y}}{\partial n_{\rm B}}   \ ,
	\label{eq:mun}
\end{eqnarray}
as well as their corresponding density derivatives
\begin{eqnarray}
	\dfrac{\partial \mu_{\rm n, MM}}{\partial n_{\rm B}} & =  & \dfrac{6}{5} \dfrac{\partial \varepsilon_{\rm F, n}^{\ast}}{\partial n_{\rm B}} + 2 \dfrac{\partial v_{\rm MM}}{\partial n_{\rm B}} + \dfrac{3}{5} n_{\rm B} \dfrac{\partial^{2} \varepsilon_{\rm F, n}^{\ast}}{\partial n_{\rm B}} + n_{\rm B} \dfrac{\partial^{2} v_{\rm MM}}{\partial n_{\rm B}} \ ,  \nonumber \\
	\dfrac{\partial \mu_{\rm n, Y}}{\partial n_{\rm B}} &=& \dfrac{6}{5} \dfrac{\partial \varepsilon_{\rm F, n}}{\partial n_{\rm B}} + 2 \dfrac{\partial v_{\rm Y}}{\partial n_{\rm B}} + \dfrac{3}{5} n_{\rm B} \dfrac{\partial^{2} \varepsilon_{\rm F, n}}{\partial n_{\rm B}} + n_{\rm B} \dfrac{\partial^{2} v_{\rm Y}}{\partial n_{\rm B}}   \ ,
	\label{eq:dmun}
\end{eqnarray}
whose expressions are written above for both the standard MM and the YGLO functional. In Eqs.~\eqref{eq:deta} and ~\eqref{eq:d2eta}, we denoted:
\begin{equation}
	\dfrac{\partial x_{\chi}^{\rm MM}}{\partial n_{\rm B}} = \dfrac{1}{n_{\rm B}^{\rm MM} - n_{\rm B}^{\chi}}    
\end{equation}
\begin{equation}
	f^{\prime} (x_{\chi}^{\rm MM}) =
	\begin{cases}  
		\dfrac{1}{\left(x_{\chi}^{\rm MM}\right)^{2}} \exp\left(-\dfrac{1}{x_{\chi}^{\rm MM}}\right), \qquad & x_{\chi}^{\rm MM} > 0 \\
		0, \qquad                                                                             & \textrm{elsewhere}    
	\end{cases}
\end{equation}
and
\begin{equation}
	f^{\prime\prime} (x_{\chi}^{\rm MM}) =
	\begin{cases}  
		\dfrac{1 - 2 x_{\chi}^{\rm MM}}{\left(x_{\chi}^{\rm MM}\right)^{4}} \exp\left(-\dfrac{1}{x_{\chi}^{\rm MM}}\right), \qquad & x_{\chi}^{\rm MM} > 0  \\
		0, \qquad                                                                             & \textrm{elsewhere} \ . 
	\end{cases}
\end{equation}
Moreover, Eqs.~\eqref{eq:mun} and 
~\eqref{eq:dmun} require to derive the second derivative with respect to the density of the Fermi energy, that is
\begin{eqnarray}
	\dfrac{\partial^{2} \varepsilon_{\rm F,q}^{\ast}}{\partial n_{\rm B}^{2}} 
	&=&  \dfrac{1}{n_{\rm B}} \left( \dfrac{\partial \varepsilon_{\rm F,q}^{\ast}}{\partial n_{\rm B}} - \dfrac{\varepsilon_{\rm F,q}^{\ast}}{n_{\rm B}} \right) \left[ \dfrac{m_{\rm q}^{\ast}}{m} \left( \kappa_{\rm sat} + \tau_{3} \kappa_{\rm sym} \delta \right) \left( \dfrac{n_{\rm B}}{n_{\rm sat}} \right)
	+ \dfrac{2}{3} \right] + \dfrac{\varepsilon_{\rm F,q}^{\ast}}{n_{\rm B}} \left(\dfrac{m_{\rm q}^{\ast}}{m} \right)^{2} \left( \dfrac{\kappa_{\rm sat} + \tau_{3} \kappa_{\rm sym} \delta }{n_{\rm sat}} \right) 
\end{eqnarray}
and the second derivatives of the potential energy per nucleon with respect to the density, that is
\begin{equation}
	\dfrac{\partial^{2} v_{\rm MM} (n_{\rm B}, \delta)}{\partial n_{\rm B}^{2}}  = \sum_{\alpha = 2}^{4} \dfrac{1}{\left(\alpha-2\right)!} \left(v_{\alpha}^{\rm is} + v_{\alpha}^{\rm iv} \delta^{2}  \right) x^{\alpha-2},  
\end{equation}
for the standard MM, and
\begin{eqnarray}
	\dfrac{\partial^{2} v_{\rm Y} (n_{\rm B}, \delta)}{\partial n_{\rm B}^{2}} 
	& = & \dfrac{\partial^{2} Y_{\rm s}}{\partial n_{\rm B}^{2}}n_{\rm B} + 2 \dfrac{\partial Y_{\rm s}}{\partial n_{\rm B}} + \dfrac{10}{9}D_{\rm s} n_{\rm B}^{-1/3} + \alpha (\alpha + 1) F_{\rm s} n_{\rm B}^{\alpha - 1} \nonumber \\
	&& + \left[ \dfrac{\partial^{2} Y_{\rm n}}{\partial n_{\rm B}^{2}}n_{\rm B} + 2 \dfrac{\partial Y_{\rm n}}{\partial n_{\rm B}} + \dfrac{10}{9}\dfrac{D_{\rm n}}{ n_{\rm B}^{1/3}} + \alpha (\alpha + 1) F_{\rm n} n_{\rm B}^{\alpha - 1} 
	- \dfrac{\partial^{2} Y_{\rm s}}{\partial n_{\rm B}^{2}}n_{\rm B} - 2 \dfrac{\partial Y_{\rm s}}{\partial n_{\rm B}} - \dfrac{10}{9}\dfrac{D_{\rm s} }{n_{\rm B}^{1/3}} - \alpha (\alpha + 1) F_{\rm s} n_{\rm B}^{\alpha - 1}  \right] \delta^{2}, \nonumber \\
\end{eqnarray}
where 
\begin{eqnarray}
	\dfrac{\partial^{2} Y_{\rm i}}{\partial n_{\rm B}^{2}}
	& = & \dfrac{2 Y_{\rm i}}{3 B_{\rm i} n_{\rm B}} \left[ \dfrac{\partial Y_{\rm i}}{\partial n_{\rm B}} \left( R_{\rm i} n_{\rm B}^{1/3} - 2C_{\rm i} n_{\rm B}^{2/3} \right) - \dfrac{Y_{\rm i}}{3 n_{\rm B}} \left( R_{\rm i} n_{\rm B}^{1/3} - C_{\rm i} n_{\rm B}^{2/3} \right) \right], \qquad {\rm i = s, n}
\end{eqnarray}
for the YGLO functional.
	
\end{widetext}

\bibliography{ref}

\end{document}